\documentclass[iop]{emulateapj_pdftex}
\usepackage{amsmath,dsnr,epsfig,ifthen,natbib}
\usepackage[caption=false]{subfig}
\newcommand{\hst}{{\it HST}}
\newcommand{\vfifty}{\ifmmode v_{50\%}\else $v_{50\%}$\fi}
\newcommand{\vfiftyave}{\ifmmode \langle v_{50\%}\rangle \else $\langle
  v_{50\%}\rangle$\fi}
\newcommand{\vtsig}{\ifmmode v_{98\%}\else $v_{98\%}$\fi}
\newcommand{\vtsigave}{\ifmmode \langle v_{98\%}\rangle \else $\langle
  v_{98\%}\rangle$\fi}
\newcommand{\weq}{\ifmmode W_{eq}\else $W_{eq}$\fi}
\newcommand{\weqa}{\ifmmode W_{eq}^{abs}\else $W_{eq}^{abs}$\fi}
\newcommand{\weqe}{\ifmmode W_{eq}^{em}\else $W_{eq}^{em}$\fi}
\newcommand{\sbe}{\ifmmode I^{em}\else $I^{em}$\fi}
\newcommand{\nh}{\ifmmode N(H)\else $N(H)$\fi}
\newcommand{\sersic}{S\'{e}rsic}

\shorttitle{Extended \nad\ Emission in a Nearby Galactic Wind}
\shortauthors{Rupke \& Veilleux}

\begin{document}

\slugcomment{Accepted to ApJ 29 Jan 2015}

\title{Spatially Extended \nad\ Resonant Emission and Absorption in
  the Galactic Wind of the Nearby Infrared-Luminous Quasar
  F05189$-$2524}

\author{David S. N. Rupke} \affil{Department of Physics, Rhodes
  College, Memphis, TN 38112} \email{drupke@gmail.com}

\author{and Sylvain Veilleux} \affil{Department of Astronomy and Joint
  Space-Science Institute, University of Maryland, College Park, MD
  20742}

\begin{abstract}
  Emission from metal resonant lines has recently emerged as a
  potentially powerful probe of the structure of galactic winds at low
  and high redshift. In this work, we present only the second example
  of spatially resolved observations of \nad\ emission from a galactic
  wind in a nearby galaxy (and the first 3D observations at any
  redshift). F05189$-$2524, a nearby ($z=0.0428$) ultraluminous
  infrared galaxy powered by a quasar, was observed with the integral
  field unit on the Gemini Multi-Object Spectrograph (GMOS) at Gemini
  South. \nad\ absorption in the system traces dusty filaments on the
  near side of an extended, AGN-driven galactic wind (with projected
  velocities up to 2000~\kms). These filaments ($A_V\la4$ and
  $\nh\la10^{22}$ cm$^{-2}$) simultaneously obscure the stellar
  continuum and \nad\ emission lines. The \nad\ emission lines serve
  as a complementary probe of the wind: they are strongest in regions
  of low foreground obscuration and extend up to the limits of the
  field of view (galactocentric radii of 3~kpc). An azimuthally
  symmetric \sersic\ model extincted by the same foreground screen as
  the stellar continuum reproduces the \nad\ emission line surface
  brightness distribution except in the inner regions of the wind,
  where some emission-line filling of absorption lines may occur. The
  presence of detectable \nad\ emission in F05189$-$2524 may be due to
  its high continuum surface brightness at the rest wavelength of
  \nad. These data uniquely constrain current models of cool gas in
  galactic winds and serve as a benchmark for future observations and
  models.
\end{abstract}



\section{INTRODUCTION} \label{sec:introduction}

Absorption-line probes of galactic winds are now a ubiquitous tool for
studying their properties \citep{veilleux05a}. The suite of metal
resonant line transitions in the optical and UV are a particularly
powerful tool for finding outflows in the nuclear spectra of galaxies
at both low $z$
\citep[e.g.,][]{heckman00a,rupke02a,rupke05a,rupke05b,martin05a} and
high $z$
\citep[e.g.,][]{shapley03a,weiner09a,rubin10a,martin12a}. Spatially-resolved
observations of these lines have recently enabled better constraints
on the properties of the cool gas phase of galactic winds in nearby
systems \citep{shih10a,rupke11a,rupke13a}.

Redshifted emission from these same transitions has been found in
stacked spectra at low and high redshift
\citep{shapley03a,weiner09a,rubin10a,chen10a,erb12a,kornei13a}, as
well as in a handful of individual spectra
\citep{martin09a,rubin11a,martin13a}. This emission is scattered into
the line of sight by the outflow due to the fact that the absorption
and re-emission occur at the same wavelength. Resonant line emission
from cool gas in galactic winds should be a readily observable feature
of these winds in the absence of significant dust or strong
collimation \citep{prochaska11a}. The properties of this emission can
be used to constrain the wind density and geometry.

However, only one case of spatially-resolved resonant emission from a
galactic wind has been found in the nearby universe
\citep{phillips93a}, despite the large number of nearby outflows that
show resonant absorption. NGC 1808 is a nearby early-type spiral
inclined 57\arcdeg\ to the line of sight with long dust filaments
extending from the nucleus \citep[][and references
therein]{phillips93a}. \citet{phillips93a} discovered that the dusty
filaments on the near side of a nuclear, bipolar outflow show
blueshifted \nad\ absorption at projected galactocentric radii of
$\la$2~kpc, while redshifted \nad\ emission lines emerge behind the
disk on the far side of the outflow.

Deep observations of two galactic winds at $z=0.7-0.9$ also show
spatially resolved \ion{Mg}{2} 2796, 2800 \AA\ emission at
$\sim$10~kpc projected radii \citep{rubin11a,martin13a}. These two
systems are blue star forming galaxies that show blueshifted
absorption and redshifted emission, similar to NGC~1808 but on larger
scales.

\begin{figure*}[t]
  \centering \includegraphics[width=6.5in]{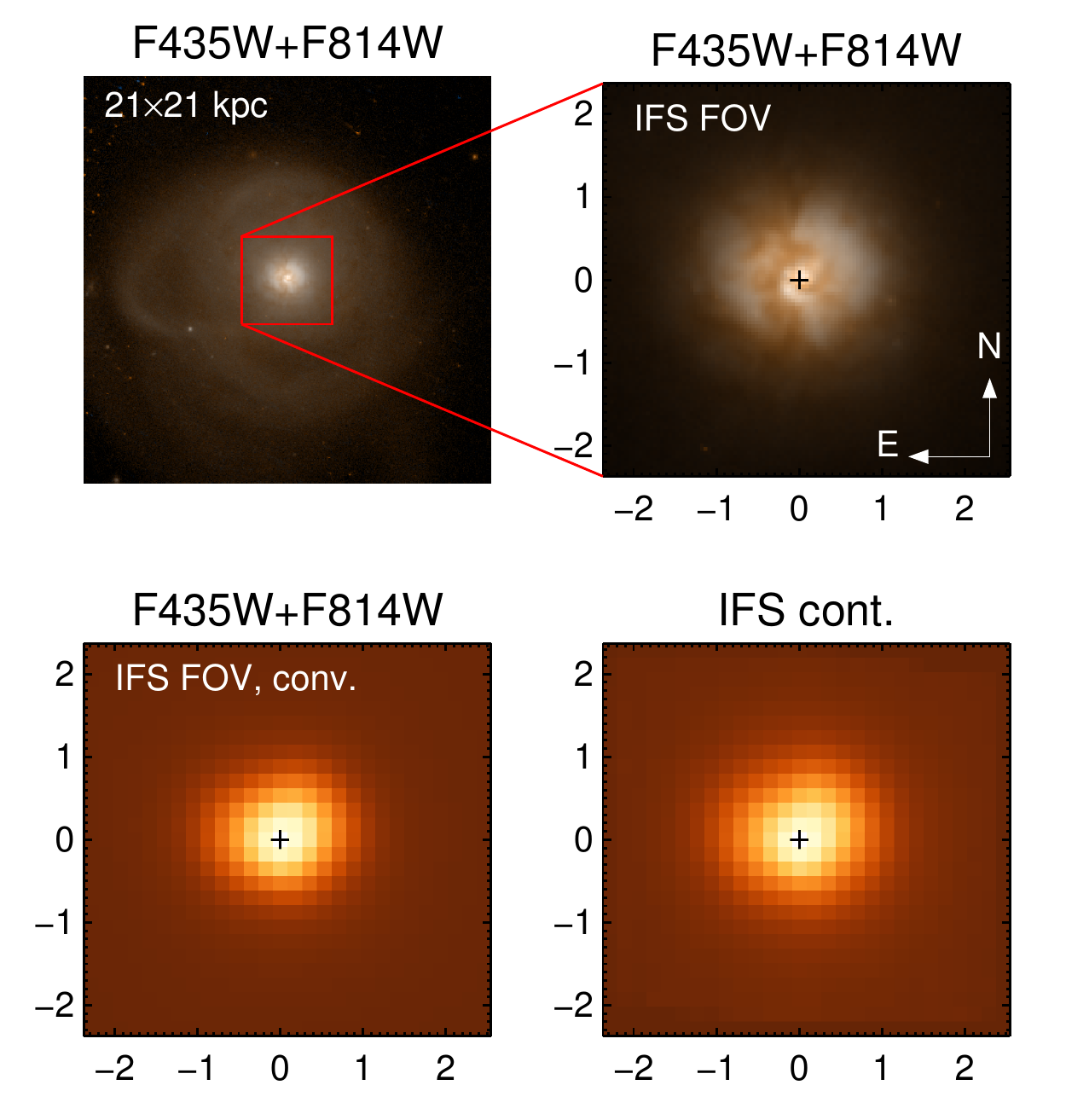}
  \caption{Continuum images of F05189$-$2524: (top left) \hst\ 3-color
    image constructed from ACS F435W and F814W exposures, FOV $=$
    25''$\times$25''; (top right) the same image zoomed into the GMOS
    FOV (5\farcs6$\times$5\farcs4); (bottom left) the average of the
    F435W and F814W exposures which is then zoomed, smoothed with a
    Gaussian kernel of FWHM $=$ 0\farcs6 to match the ground-based
    seeing, and rebinned to approximate the IFS data; and (bottom
    right) the GMOS data, summed between 5600 and 6400~\AA. The IFS
    continuum peak is denoted by the cross, and the axes give
    galactocentric coordinates in kpc. The top panel scaling is asinh
    \citep{lupton99a} with $\beta=0.05$ (left) and $\beta=1$ (right);
    the lower panel scales are linear.}
  \label{fig:map_cont}
\end{figure*}

\begin{figure*}[t]
  \centering \includegraphics[width=6in]{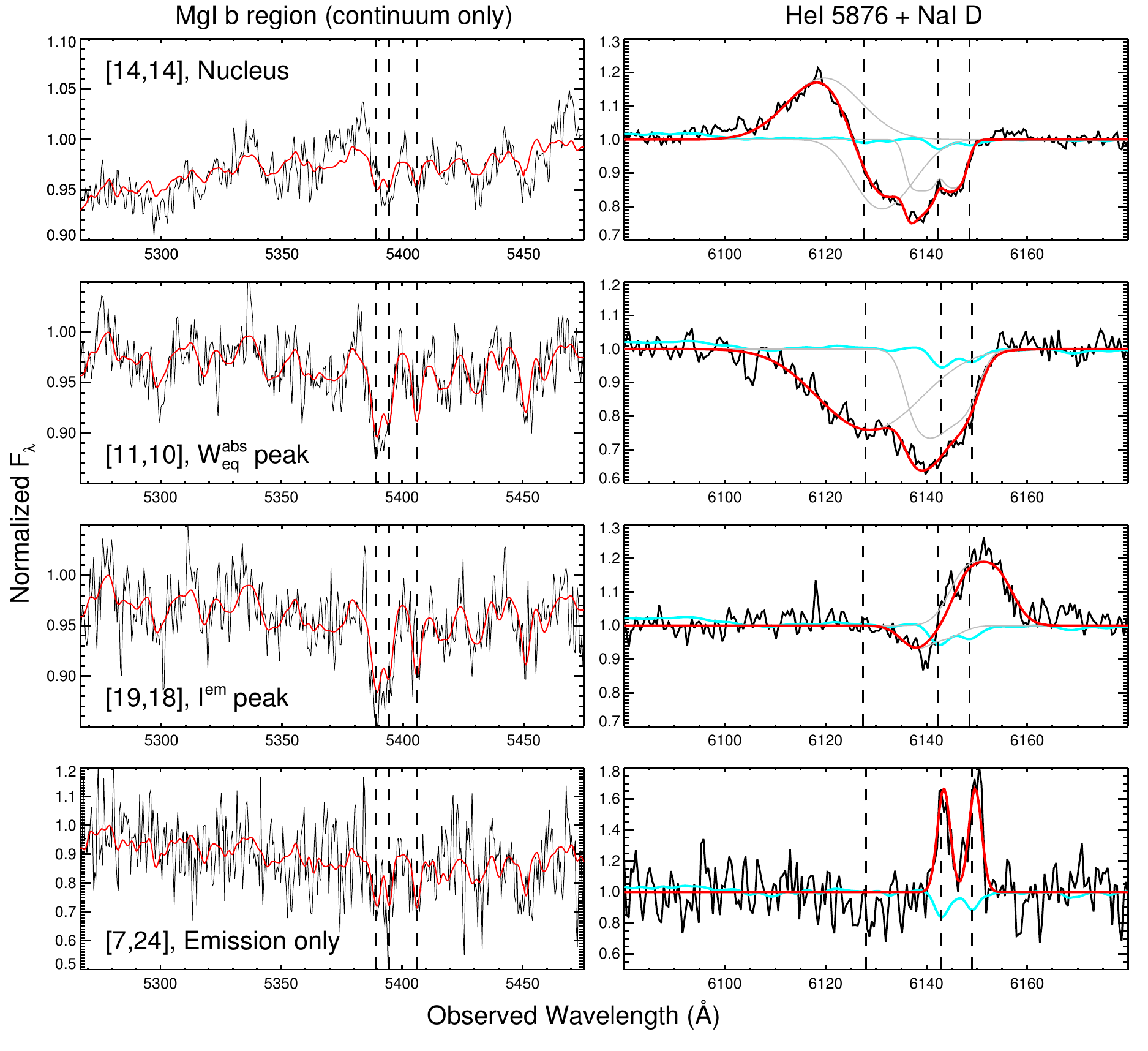}
  \caption{Examples of fits. (Left) Continuum fits to the
    \ion{Mg}{1}~b region. The narrow [\ion{N}{1}] 5198, 5200~\AA\ and
    broad [\ion{Fe}{7}] 5159~\AA\ emission lines have been fit and
    removed. Data is in solid black and total fits are in red. The
    wavelengths of the \ion{Mg}{1}~b triplet as expected from the
    stellar redshift in each spaxel are shown as vertical dashed
    lines. Each panel is labeled with the spaxel column and row number
    and any relevant \nad\ properties. The top panel fit is affected
    by residuals from strong and broad [\ion{Fe}{7}] emission. (Right)
    Fits to \nad\ and \ion{He}{1} 5876~\AA. The data is in solid
    black, the fits to individual components are in gray, the total
    fits are in red, and the expected wavelengths of \nad\ and
    \ion{He}{1}~5876~\AA\ are shown as vertical dashed lines. The
    underlying stellar continuum by which the data was normalized is
    shown as solid cyan; the correction for stellar \nad\ absorption
    is evidently small.}
  \label{fig:example_fits}
\end{figure*}

The nearby ($z = 0.0428$) galaxy F05189$-$2524 is an ultraluminous
infrared galaxy (ULIRG) with a Seyfert 2 optical spectral type
\citep{veilleux99a}. It is one of the nearest and brightest ULIRGs,
and reveals a hidden broad line region in the near-IR
\citep{veilleux99b}. Approximately 70\%\ of the bolometric luminosity
of F05189$-$2524 arises from an AGN \citep{veilleux09a}, meaning that
the AGN in this system exceeds the quasar threshold in luminosity
($\sim$$10^{45}$
erg s$^{-1}$).
It has a high-velocity outflow, as seen in neutral, ionized, and
molecular gas
\citep{rupke05c,westmoquette12a,veilleux13a,teng13a,bellocchi13a}.
F05189$-$2524
was observed as part of an integral field spectroscopic survey of
nearby major mergers and quasars to study AGN feedback
(\citealt{rupke11a,rupke13a}; Rupke et al. 2015, in prep.).

\citet{rupke05c} reported that this galaxy shows extended \nad\
emission in long-slit spectra. Here we present integral field
observations that show the distribution of resonant line emission and
absorption in this system. This is only the second known example of
such emission in the local universe, and as such serves as a benchmark
for models and observations, both past and future.

In Section \ref{sec:obs} we present the observations and our methods
of data reduction and analysis. We discuss the equivalent width, \nad\
emission line flux, and velocity measurements in
Section\,\ref{sec:results}. We connect the dust obscuration of the
stellar continuum to both the column of neutral Na and the emission
line flux, and introduce possible wind structures to explain the \nad\
emission. In Section\,\ref{sec:discussion} we outline our preferred
model for the \nad\ emission and discuss emission line filling, as
well as considering the detectability of \nad\ emission in galactic
winds. We summarize in Section\,\ref{sec:summary}. Throughout the
paper we assume $H_0 = 73$~\kms~Mpc$^{-1}$, $\Omega_m = 0.27$, and
$\Omega_\Lambda = 0.73$, yielding $0.88$~kpc~arcsec$^{-1}$ at the
redshift of F05189$-$2524 ($z = 0.0428$).

\section{OBSERVATIONS, REDUCTION, AND ANALYSIS} \label{sec:obs}

F05189$-$2524 was observed with the Integral Field Unit in the Gemini
Multi-Object Spectrograph (GMOS; \citealt{allington-smith02a,hook04a})
on the Gemini South telescope. The integral field unit was used in
1-slit mode with the B600 grating, yielding wavelength coverage from
4600 to 7400~\AA\ and a spectral resolution of 1.6~\AA\ at
6000~\AA. The Gemini IRAF package (v1.12) and IFSRED\footnote{\tt
  http://github.com/drupke/ifsred} \citep{rupke14a} was employed to
reduce the data. \citealt{rupke13a} present the details of the data
reduction. The sky was subtracted using the average sky spectrum from
the 250 spaxels offset 1\arcmin\ from the field of view (FOV). The
standard deviation in the sky flux across the FOV was $<$4\%\ and the
median sky fluxes in the sky and science spaxels differed by $<$4\%,
as determined from fits to the [\ion{O}{1}] 5577~\AA\ sky line. Seven
dithered 30-minute science exposures were combined to yield a FOV of
5\farcs6$\times$5\farcs4, and the data was resampled to square spaxels
of side 0\farcs2. The median seeing was 0\farcs6.

Figure~\ref{fig:map_cont} shows {\em Hubble Space Telescope} (\hst)
Advanced Camera for Surveys (ACS) images constructed from F435W and
F814W exposures \citep{armus09a,kim13a}. It also presents the IFS
continuum integrated between 5600 and 6400~\AA, compared to a smoothed
and rebinned average of the F435W and F814W images.

The continuum and emission lines in each spaxel were modeled using
IFSFIT\footnote{\tt http://github.com/drupke/ifsfit}
\citep{rupke14b}. In brief, IFSFIT masks emission line regions, fits
the continuum using PPXF \citep{cappellari04a}, and then
simultaneously fits all emission lines in the continuum-subtracted
spectrum. PPXF yields an accurate stellar redshift in each spaxel
based on the entire continuum. The range $4600-7100$~\AA\ was fit,
which contained the absorption and emission lines of interest. A
linear combination of stellar templates from
\citet{gonzalezdelgado05a} and additive Legendre polynomials up to
fourth degree yielded a good fit to the continuum of
F05189$-$2524. Though the continuum fit parameters may not be unique
due to degeneracies in the continuum models, the fitting nevertheless
does an excellent job of reproducing the stellar absorption lines in
the continuum and the continuum shape, which is its sole purpose.
Figure~\ref{fig:example_fits} illustrates the quality of the fits
using the continuum region near \ion{Mg}{1}~b, since the strengths of
stellar \ion{Mg}{1}~b and stellar \ion{Na}{1}~D are proportional
\citep[see][and references therein]{rupke02a}. The fits are of high
quality even in regions strongly affected by emission line residuals
(the top left panel of Figure~\ref{fig:example_fits}, which has had a
strong, broad [\ion{Fe}{7}] 5159~\AA\ line removed) or that are noisy
(the bottom left panel).

\begin{figure*}
  \centering \includegraphics[width=6in]{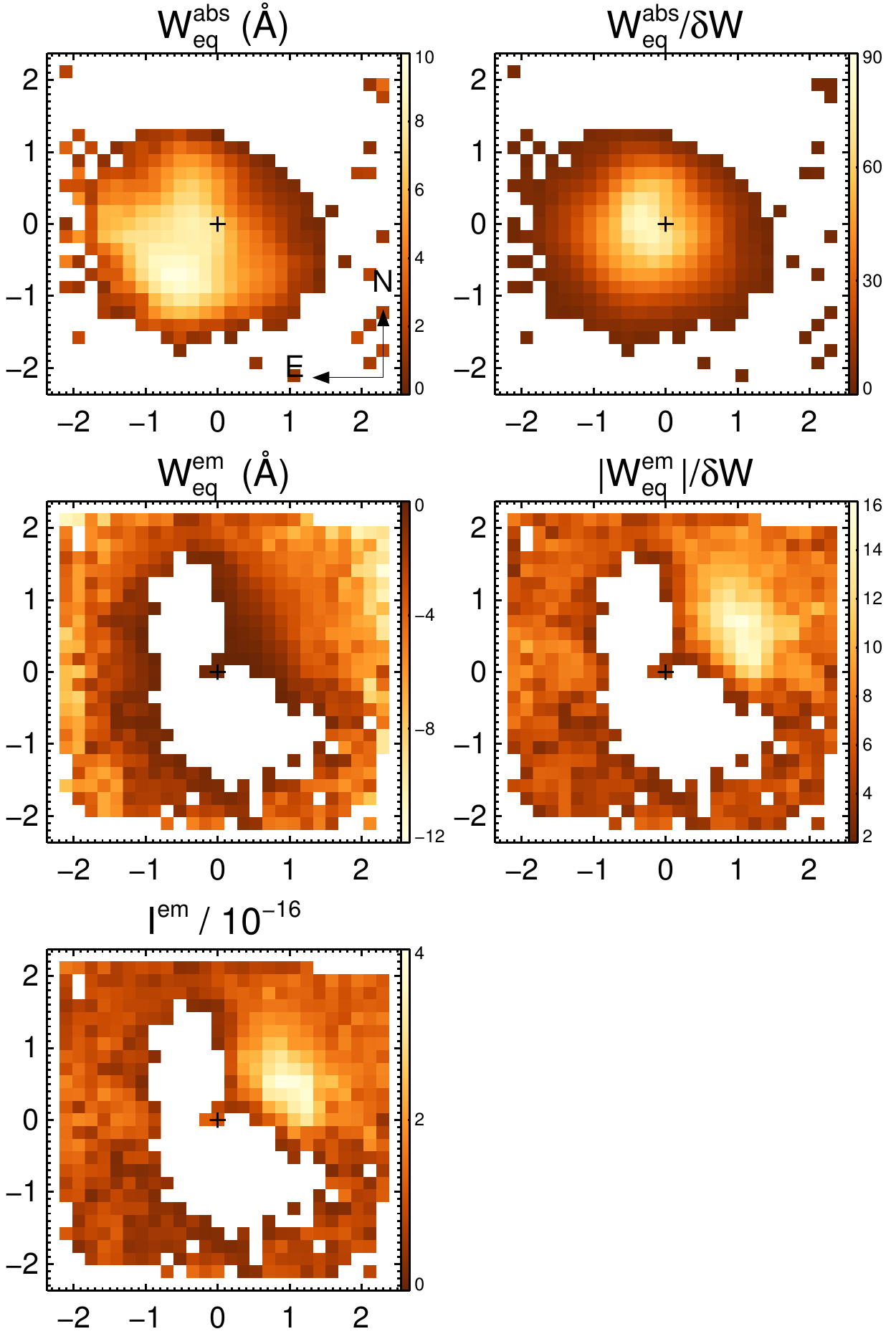}
  \caption{(Top) Observed equivalent width (\AA) of \nad\ absorption
    and its signal-to-noise ratio (SNR). Absorption lines were
    detected and integrated using an empirical method and corrected
    for contamination by the nuclear PSF (\S\ref{sec:obs}). A
    3$\sigma$ threshold was used to mask the data. (Middle) Observed
    equivalent width (\AA) of \nad\ emission and its SNR. (Bottom)
    Surface brightness of \nad\ emission (units of $10^{-16}$ erg
    s$^{-1}$ cm$^{-2}$ arcsec$^{-2}$). The values shown in this figure
    are the combined measurements of both doublet lines.}
  \label{fig:map_empweq}
\end{figure*}

Before modeling the emission lines, the stellar$+$polynomial continuum
was subtracted. Emission lines were modeled with 1--3 Gaussian
velocity components. Further details of the emission-line fits for
this object will be discussed in a forthcoming paper on outflows in a
sample of quasars (Rupke et al. 2015, in prep.).

The \nad\ doublet in this galaxy shows absorption, emission, or both
across the entire FOV. An empirical and a fit-based approach were
employed to measure the properties of the lines. Prior to modeling
\nad, the spectrum was divided by the continuum fit. This removed the
stellar contribution to \nad, which has a median and standard
deviation of $W_{eq} = 0.9\pm0.6$~\AA\ based on our fits. This stellar
contribution is much smaller than the interstellar contribution except
at the lowest measured equivalent widths. At these levels the
interstellar line is typically strongly blue- or redshifted from the
stellar line (\S\,\ref{sec:disc_velocity}). A second-order
multiplicative polynomial was then fit to the region surrounding the
\nad -- \ion{He}{1} 5876~\AA\ complex to account for any local
systematic errors in the continuum fitting. In
Figure~\ref{fig:example_fits} we plot both the normalized data and the
continuum fit near \nad\ to illustrate that the normalization and
removal of the stellar absorption lines change the spectra minimally,
and accordingly account for negligible error in the \nad\ fits.

Example spectra illustrating the range of fit types (described below)
are shown in Figure~\ref{fig:example_fits}. Errors in fit parameters
and equivalent widths were determined using Monte Carlo methods
\citep{rupke05a}.

For the empirical method, the data (signal) and error (noise) spectra
were boxcar smoothed using a kernel with FWHM equal to the spectral
resolution \citep{cooksey08a}. Regions where $|signal/noise| > 1.6$
were identified as absorption or emission line detections; the
threshold was determined by visual examination of the detection
boundaries. The resulting maps of equivalent widths, signal-to-noise
ratios (SNRs), and emission-line surface brightnesses (all from direct
integration) are shown in Figure~\ref{fig:map_empweq}. These values
all refer to the combined measurements of both doublet lines.

The fitting method required careful treatment of the emission and
absorption lines. The emission lines were assumed to each consist of a
single, Gaussian velocity component. The flux ratio in the doublet
lines can vary between the optically thick
($I^{em}_{5890}/I^{em}_{5896} = 1$) and thin limits
($I^{em}_{5890}/I^{em}_{5896} = 2.01$, equal to the optical depth
ratio of the two lines). However, the line ratio is difficult to
independently constrain in many spaxels. In spaxels with the highest
SNR, the lines are broad and fits with a given line ratio are
statistically indistinguishable from others due to the close
separation of the two lines (304~\kms). In spaxels with
low-to-moderate SNR and adjacent absorption, the chance of absorption
mixing with line emission could lead to an underestimate of the flux
ratio, since the mixing would preferentially lower the SNR of the
5890~\AA\ line (see \S\,\ref{sec:disc_structure} for more discussion
of emission line filling).

The best spaxels for determining the intrinsic line flux ratio are
therefore those with narrow line widths and pure emission. In the
current data, these regions occur primarily in the northeast and
northwest regions of the FOV. In the NE and NW, two regions were
identified, each of $25-26$ adjacent spaxels which had
well-constrained emission line ratios. The median best-fit $\sigma$ in
these spaxels is 62~\kms, and the median emission-line SNR of the
equivalent width is 10.5. The median best-fit line ratio is 1.00, and
Monte Carlo simulations of the errors yielded a median 68\%\ (95\%)
confidence level of $I^{em}_{5890}/I^{em}_{5896} < 1.14 (1.55)$. Based
on these results, the flux ratio for all other spaxels was fixed to
the optically thick case ($I^{em}_{5890}/I^{em}_{5896} = 1$).

\begin{figure*}[t]
  \centering \subfloat{{\includegraphics[width=3.5in]{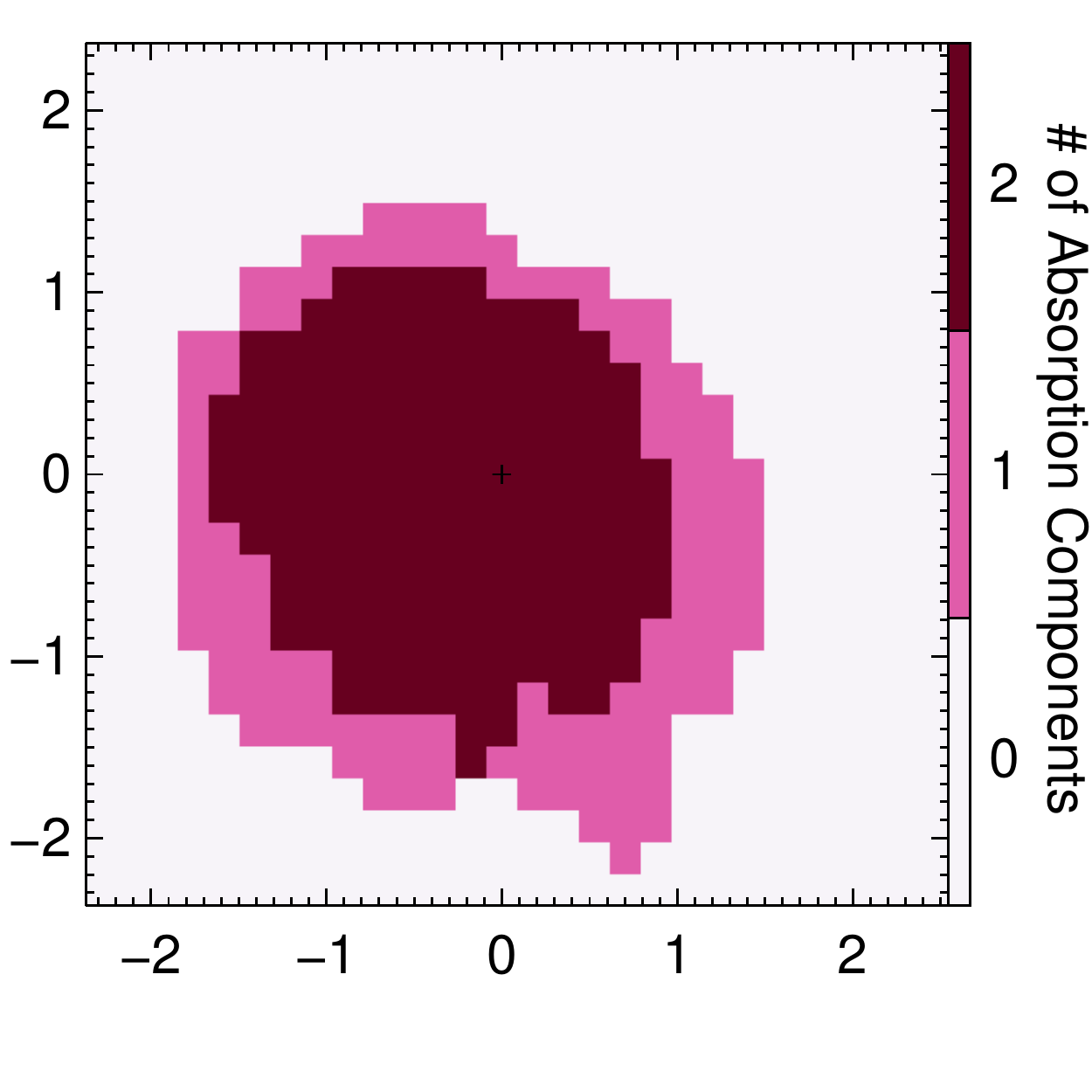} }}%
  \subfloat[]{{\includegraphics[width=3.5in]{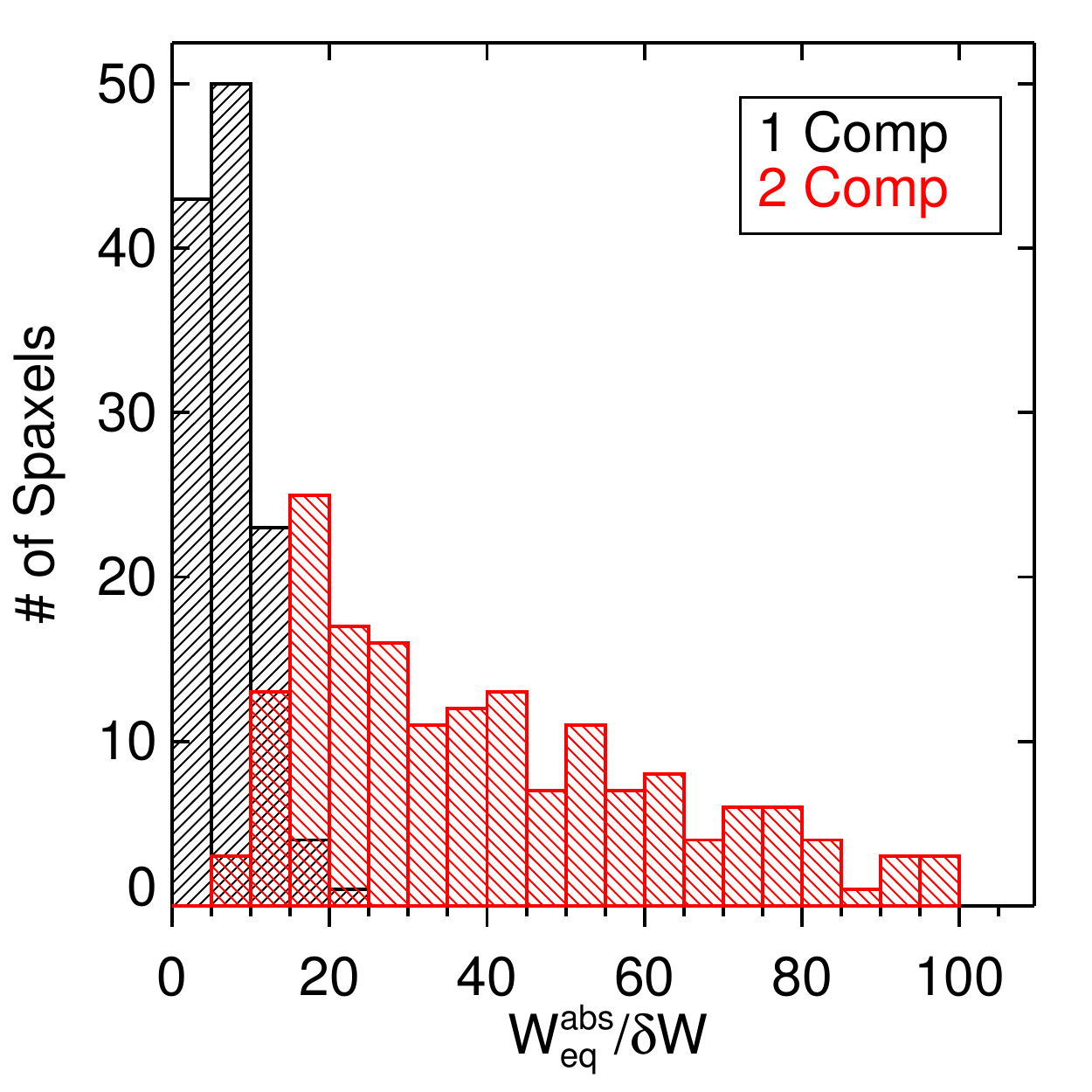} }}%
  \caption{(Left) Map of number of components in the \nad\ absorption
    line fit. The number of components that can be fit is primarily a
    function of the SNR, which results in the number of components
    decreasing with galactocentric distance. (Right) Histograms of
    absorption line SNR for one-component (black hatched) and
    two-component (red hatched) fits.}
  \label{fig:ncomp}
\end{figure*}

The absorption line fitting followed \citet{rupke05a}, which assumes a
Gaussian in optical depth and allows for a non-zero (but constant)
covering factor in each velocity component. The absorption lines in
F05189$-$2524 clearly require two velocity components in most of the
spaxels with $\text{SNR}\ga15$ (Figure \ref{fig:ncomp}). In spaxels
with lower SNR, two components did not yield a well-constrained fit,
and only one component was used.

The fits tended toward one of two solutions: an optically thick
($\tau_{5890}\sim2$) fit with covering factor $C_f\sim0.2$, and an
optically thin ($\tau_{5890}\sim0.1$) fit with $C_f=1.0$. The
one-component fits (which also had lower SNR) tended to yield lower
total optical depths than the two-component fits. We argue in Section
\,\ref{sec:weq_flux} that the high-$\tau$ fit is generally the correct
one.

Accurately measuring the properties of the \nad\ absorption on the
blue edge of the feature requires accurate modeling of the \ion{He}{1}
5876~\AA\ emission line in spaxels where it is present. Emission line
fits to \ion{He}{1} 6678~\AA\ were used to constrain the properties of
\ion{He}{1} 5876~\AA\ emission. \ion{He}{1} lines in F05189$-$2524 are
broad and blueshifted, and the 6678~\AA\ line was fit with a single
Gaussian in the central spaxels (it was too weak at larger
galactocentric radii). The central wavelength and line width of the
5876~\AA\ line were fixed to those of 6678~\AA, while its peak flux
was allowed to vary. Detected \ion{He}{1} 5876~\AA\ emission is
confined to radii of three spaxels or less.

In $\sim$30\%\ of spaxels containing the \nad\ line, both absorption
and emission are obviously present. The two are adjacent in wavelength
space in these spaxels, with the absorption line blueshifted from the
stellar continuum and the emission line near systemic or
redshifted. Modeling the absorption and emission lines simultaneously
leads to significant degeneracy in the best fit, since increasing the
emission line flux and shifting it to the blue can be compensated for
in the fit by increasing the absorption line equivalent width and
shifting it to the red. To model the \nad\ in these spaxels, the
emission and absorption lines were thus fit separately. The feature
boundaries determined from the empirical method were used to first fit
the emission line only using data redward of the absorption line. The
emission line properties were then fixed in the absorption line fit.

As a result of this method, it is possible that the emission line
equivalent widths are underestimated and that their measured line
centers are shifted redward of the true line centers. Correspondingly,
the absorption line equivalent widths could be underestimated and the
measured line centers shifted blueward. This effect is explored in
\S\S\,\ref{sec:disc_velocity} and \,\ref{sec:disc_structure}.

\begin{figure*}
  \centering \includegraphics[width=5.75in]{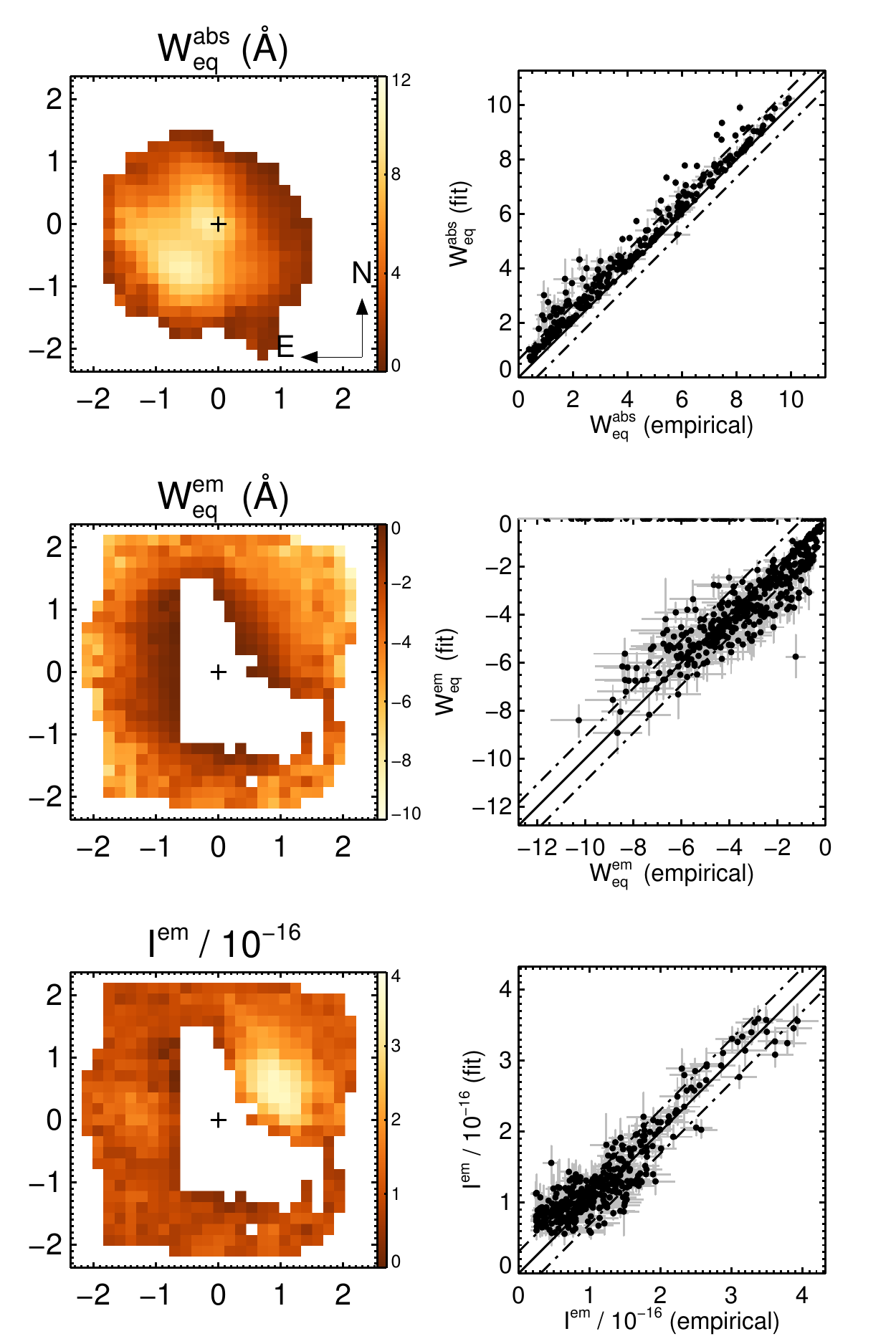}
  \caption{(Top) Observed equivalent width (\AA) of \nad\ absorption
    from fitting with analytic profiles (\S\ref{sec:obs}), and a
    comparison of empirical and fit-based equivalent widths. The solid
    line is equality, and the dot-dashed lines represent the RMS
    scatter. (Middle) \nad\ emission line equivalent width (\AA) and
    comparison of methods. (Bottom) \nad\ emission line surface
    brightness (units of $10^{-16}$ erg s$^{-1}$ cm$^{-2}$
    arcsec$^{-2}$) and comparison of methods. The values shown in this
    figure are the combined measurements of both doublet lines.}
  \label{fig:map_fitweq}
\end{figure*}

The maps of fit-based equivalent widths and emission-line surface
brightnesses, as well as comparisons of the two methods, are shown in
Figure~\ref{fig:map_fitweq}. There is excellent correspondence between
the empirical and fitting methods. The fitting method allows better
deblending of emission and absorption, and as a result yields slightly
higher absorption line equivalent widths (typically $\la0.5\AA$,
though the difference is higher near the nucleus where \ion{He}{1}
5876~\AA\ is strong).

F05189$-$2524 contains an unresolved nuclear point source due to an
AGN \citep{surace98a,kim02a,farrah05a,veilleux06a,kim13a}, which
contributes $<$1\%\ of the galaxy's light \citep{kim13a}. The additive
polynomials in our continuum fit may partly trace light from this
point source, as the polynomial dominates within a few$\times$0\farcs1
of the nucleus. However, the shape of the intrinsic AGN continuum is
unknown and the polynomial-only surface brightness profile is broader
than the seeing disk. To more reliably model the spatially unresolved
AGN emission, the PSF of the data cube was traced using the bright
[\ion{Fe}{7}] 6087~\AA\ line (which is unresolved even at \hst\
resolution; \citealt{farrah05a}). The AGN-only continuum emission was
then assumed to take the shape of this PSF at the wavelength of \nad\
and to contribute 50\%\ of the total (or AGN + stellar) continuum
surface brightness at the PSF center. The normalization of the AGN
contribution is uncertain, but is broadly consistent with the more
detailed photometric modeling of \citet{kim13a} at \hst\ resolution.

Measured equivalent widths were corrected for the contribution of the
AGN by dividing \weq\ by the fractional contribution of the stellar
(or total $-$ AGN) continuum to the total \nad\ continuum in each
spaxel, as predicted by the PSF model. Low-resolution \hst\ spectra
suggest that the nuclear spectrum has little or no \nad\ absorption
present \citep{farrah05a}, so correcting for the AGN only affects the
continuum level and not the \nad\ line shape. The resulting correction
enhanced the nuclear values of the absorption line equivalent
width. Significant PSF corrections were confined to the inner few
spaxels of the absorption line maps, and have a negligible
quantitative impact on the results inferred from these maps.

\section{RESULTS} \label{sec:results}

This section presents physical quantities inferred from the \nad\
fits, including gas column density and velocity. It also quantifies
the connection of the absorbing and emitting gas to optical
obscuration and compares the emission line velocities to other profile
properties. The mass outflow rate and energetics of the wind will be
presented in a future paper in the context of a larger sample of
quasars with outflows (Rupke et al. 2015, in prep.).

\begin{figure*}[t]
  \centering \includegraphics[width=6.5in]{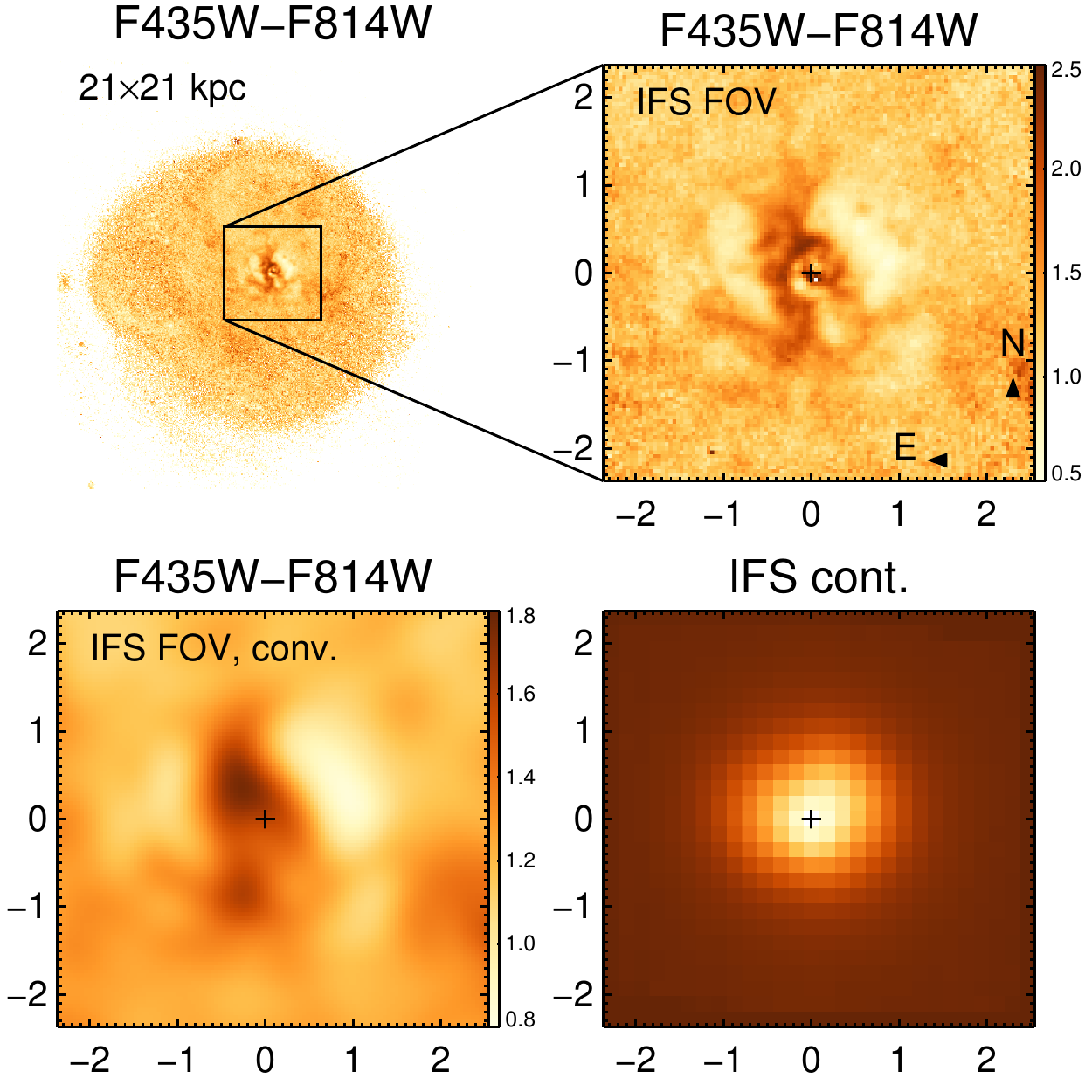}
  \caption{Color images of F05189$-$2524: (top left) $(B-I)_{AB}$
    color map constructed from F435W and F814W exposures, FOV $=$
    25''$\times$25''; (top right) the same image zoomed into the GMOS
    FOV (5\farcs6$\times$5\farcs4); (bottom left) the same image,
    zoomed and smoothed with a Gaussian kernel of FWHM $=$ 0\farcs6;
    and (bottom right) the GMOS data, summed between 5600 and
    6400~\AA.}
  \label{fig:map_col}
\end{figure*}

\begin{figure*}
  \centering \includegraphics[width=7in]{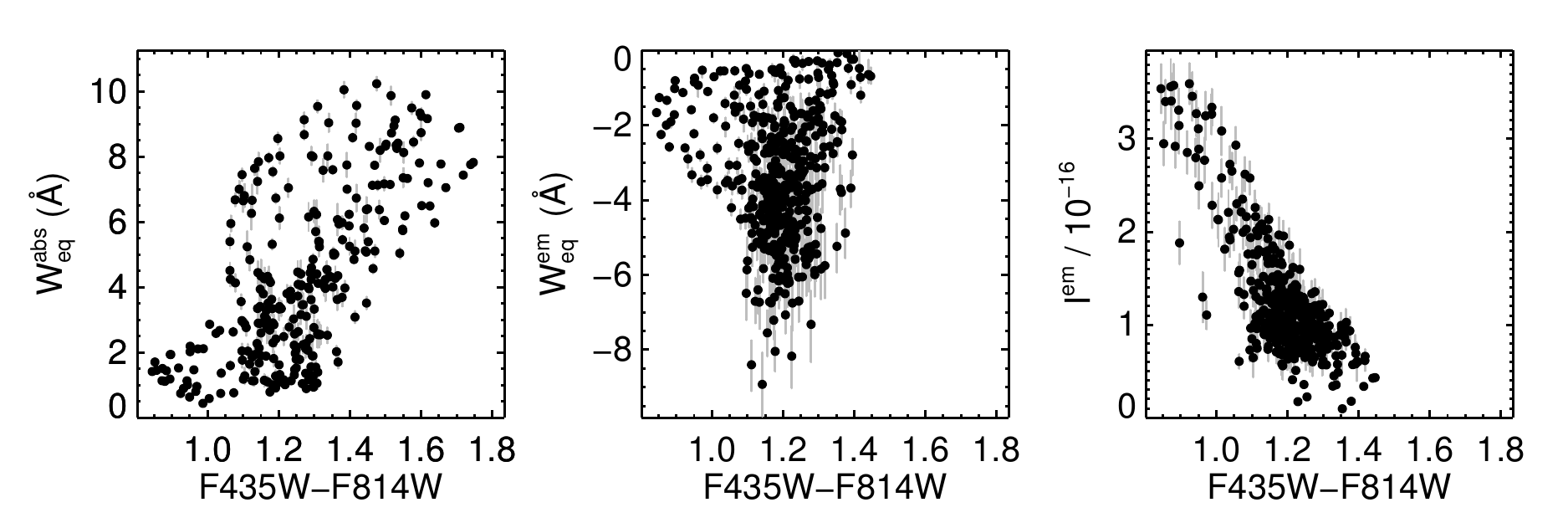}
  \caption{Observed equivalent width (\AA) of \nad\ absorption,
    equivalent width (\AA) of \nad\ emission, and surface brightness
    of \nad\ emission (units of 10$^{-16}$ erg s$^{-1}$ cm$^{-2}$
    arcsec$^{-2}$) vs. F435W$-$F814W (AB) color. The \hst\ continuum
    data was convolved to the ground-based spatial resolution and
    rebinned to match the GMOS spaxel locations and sizes. The
    apparent correlation between \weqa\ and color suggests a
    connection between dust and cool, neutral gas. The correlation
    between \sbe\ and color (but not \weqe\ and color) suggests that
    dust regulates the flux (\S\,\ref{sec:weq_flux}).}
  \label{fig:weq_v_col}
\end{figure*}

\subsection{Dusty Absorption and Obscured
  Emission} \label{sec:weq_flux}

Correlations exist between cool gas absorption and dust attenuation in
nuclear spectra of galaxies \citep{veilleux95a,heckman00a} as well as
within individual galaxies \citep{shih10a,rupke13a}, tracking the
correlations seen in interstellar sightlines in the Milky Way
\citep{hobbs74a,munari97a,vergely10a}. Filametary dust structures in
fact cover the nuclear region of F05189$-$2524 in optical images,
causing variation in the emergent optical continuum
colors. Figure~\ref{fig:map_col} shows \hst\ ACS $(B-I)$ color maps
constructed from F814W and F435W exposures (see \citealt{surace98a}
for similar WFPC2 color maps). The \hst\ data that is convolved to the
ground-based spatial resolution and rebinned to match the GMOS spaxel
locations and sizes show a range in $F435W-F814W$ (AB) colors of
$0.8-1.8$. The higher resolution data show a slightly larger range, of
$0.6-2.5$, with the upper limit occurring near the nucleus and the
lower limit in the blue region 1~kpc W of the nucleus.

The intrinsic optical spectrum of F05189$-$2524 is likely dominated by
a young stellar population. The presence of a large, concentrated
molecular gas reservoir in F05189$-$2524 \citep{sanders91a} and a
quasar-like luminosity dominated by reprocessed emission \citep{u12a}
indirectly indicate dust-enshrouded, ongoing star formation
\citep{armus07a,veilleux09a}. However, the stellar populations that
dominate the optical spectrum of F05189$-$2524 have been constrained
only with broadband photometric data \citep{surace00a}, leaving
degeneracies related to the presence of dust, metallicity, and stellar
population history \citep[e.g.,][]{calzetti01a}. The metallicity of
starburst-dominated ULIRGs is near solar \citep{rupke08a}, removing
one of these degeneracies. The stellar population modeling used in
this work (\S\ref{sec:obs}) is not comprehensive enough to precisely
resolve the remaining degeneracies, despite its ability to accurately
remove the stellar continuum. Furthermore, the GMOS spectra have a
limited wavelength range, including only two Balmer lines that are
heavily mixed with emission. Nonetheless, the stellar population age
that dominates the luminosity of most spaxels is of order several 10s
to 100s of Myr. Comprehensive modeling of large samples of ULIRGs show
that this is the typical age that dominates ULIRG optical spectra
\citep{rodriguezzaurin09a,rodriguezzaurin10a}.

The intrinsic stellar population in F05189$-$2524 should also vary
little across the FOV. The stellar populations in ULIRGs do not change
significantly within the inner 3 kpc radius
\citep{soto10a,rodriguezzaurin09a}, the region probed by our
observations. The ages and normalizations of the stellar populations
in fits to the GMOS data show the same type of population mix across
almost the entire FOV. Furthermore, the filamentary structure in the
color map (Figure \ref{fig:map_col}) is more consistent with
variations in dust column than stellar population.

We thus make the assumption that the $(B-I)$ colors in Figure
\ref{fig:map_col} represent varying levels of dust attenuation rather
than significant changes in stellar population. To estimate instrinsic
stellar colors, we used Starburst99 \citep{leitherer99a} to model the
underlying stellar population at optical wavelengths as a solar
metallicity continuous burst. Such a burst has $(B-I)_{AB} = 0.0$
after $2\times10^7$~yr. This decreases to $(B-I)_{AB} = -0.3$ at
$1\times10^7$~yr; earlier ages should not dominate the optical
spectrum \citep{rodriguezzaurin09a,rodriguezzaurin10a}. These values
bracket a small range compared to the observed variation in colors. In
our analysis we assume $(B-I)_{AB} = 0.0$; lower values will raise the
inferred $A_V$ by a small amount but otherwise do not affect our
analyses.

Using the starburst attenuation curve of \citet{calzetti00a} with
$R_V = 4.05$, the \hst\ color ranges then correspond to
$A_V = 1.2-2.8$ (convolved and rebinned) and $0.9-3.9$ (unsmoothed),
which are typical of ULIRG nuclei \citep{veilleux99a}. These $A_V$
correspond to optical depths of $1-4$ at the rest wavelength of \nad.
Figure~\ref{fig:weq_v_col} therefore shows that optical continuum
attenuation is strongly anti-correlated with \nad\ emission line
surface brightness (\sbe) and positively correlated with \nad\
absorption line equivalent width (\weqa). It has no correlation with
emission line equivalent width (\weqe).

\begin{figure}[b]
  \centering \includegraphics[width=3.5in]{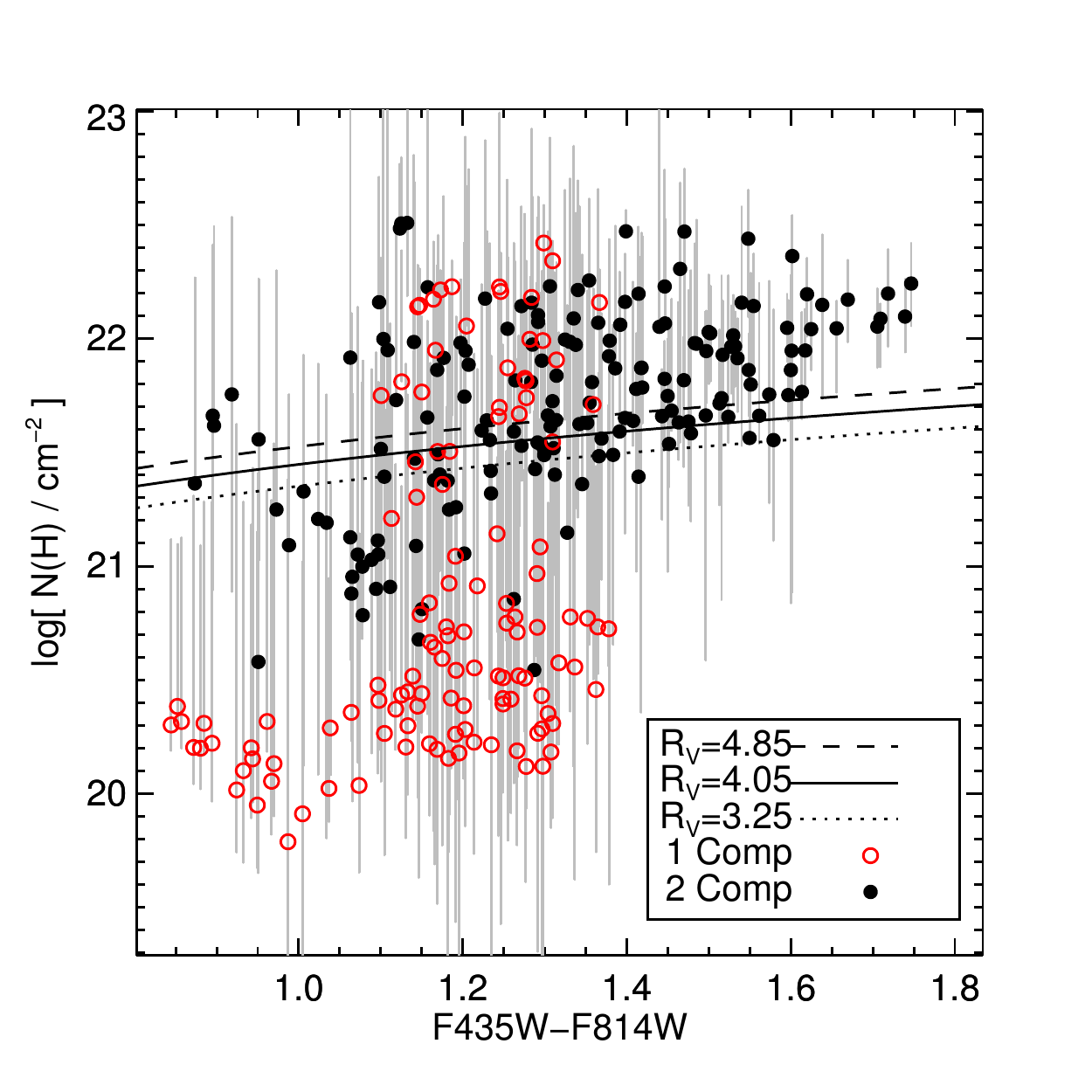}
  \caption{Column density of hydrogen vs. F435W$-$F814W color. $\nh$
    is computed from \nad\ absorption (\S\,\ref{sec:obs}). The black
    solid (red open) points represent two (one) component fits. The
    lines assume that the stellar color primarily reflects dust
    attenuation, and this fact is used to estimate \nh\ from the color
    given a plausible stellar population, a starburst attenuation
    curve \citep{calzetti00a}, and a Galactic dust-to-gas ratio
    \citep{predehl95a} (\S\,\ref{sec:disc_structure}). The discrepancy
    between one and two component fits reflects the inability of the
    one component, lower SNR spaxels to simultaneously constrain the
    covering factor and optical depth.}
  \label{fig:nh_v_col}
\end{figure}

\begin{figure*}[t]
  \centering \includegraphics[width=6.5in]{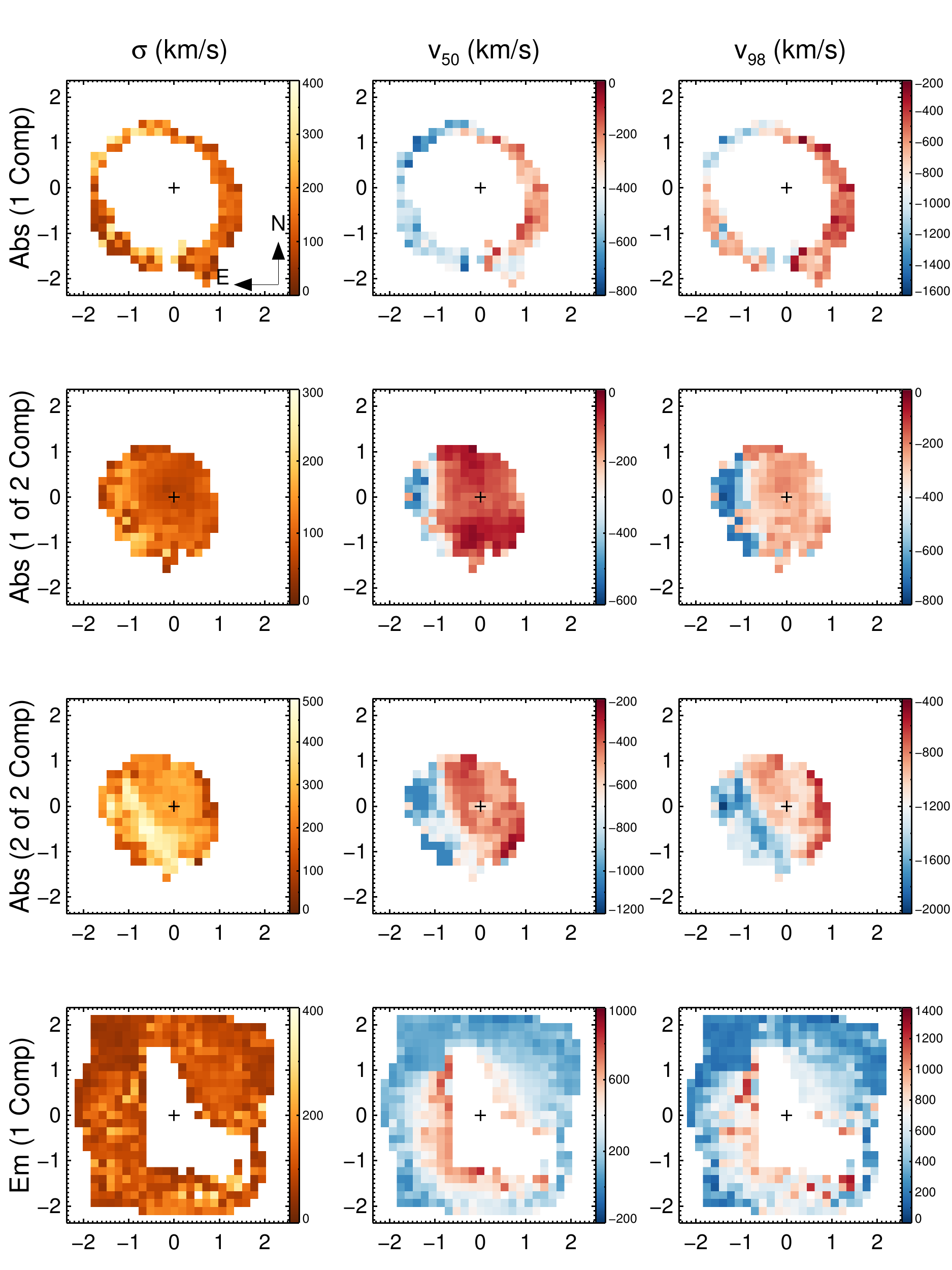}
  \caption{\scriptsize Maps of velocity dispersion (left), \vfifty\
    (middle), and $\vtsig \equiv \vfifty \pm 2\sigma$ (right) in fits
    to \nad. Absorption line fits with one (top) or two (middle)
    velocity components are shown, as well as emission line fits with
    one component (bottom). \vfifty\ and \vtsig\ in the absorption
    lines peak in a ridge running east to south, 1 kpc in projection
    from the nucleus, at blueshifted velocities up to
    $-$2000~\kms. The emission line velocities, in contrast, decrease
    with increasing galactocentric radius
    (\S\,\ref{sec:disc_velocity}).}
  \label{fig:velmaps}
\end{figure*}

However, \weqa\ is an empirical quantity that does not have a
one-to-one relationship with column density, particularly in the
presence of partial covering of the absorbing gas. The analytic fits
to \nad\ absorption accurately separate covering factor and optical
depth and allow the calculation of hydrogen column density
\citep{rupke05a}.

The ionization and dust conditions needed to compute \nh\ are assumed
to be similar to sightlines through the Milky Way \citep{rupke05a}:
ionization fraction $y\equiv 1 -
N($\ion{Na}{1}$)/N(\mathrm{Na})
= 0.9$
and dust depletion of $-0.95$~dex. The applicability of these values
to the potentially more extreme conditions of ULIRGs is
uncertain. However, a direct, independent check on \nh\ in the outflow
comes from the \ion{H}{1} absorption measurements of \citet{teng13a},
who find two outflowing components at the same velocities detected in
\nad\ (\S\,\ref{sec:disc_velocity}). For these components,
$\nh = 10^{22}$~$cm^{-2}$, which is the same as the typical column
inferred from \nad\ in the dustiest regions of this galaxy (Figure
\ref{fig:nh_v_col}). Indirect support for the validity of these
physical conditions comes from the fact that mass outflow rates
calculated using such conditions are comparable to the star formation
rate \citep{rupke05b,rupke13a} and are in agreement with estimates
from other cool gas phases \citep[e.g.,
molecular;][]{feruglio10a,sturm11a}.

The column computed from \nad\ can also be compared to the dust
column. If the varying $(B-I)$ colors represent dust attenuation, the
column density of hydrogen that is along the line of sight in each
spaxel (and is associated with the continuum-absorbing dust) can be
inferred by applying a dust-to-gas ratio to the attenuation calculated
from the stellar color. Figure \ref{fig:nh_v_col} shows that, within
the measurement uncertainties, the two-component fits do a remarkable
job of matching the dusty gas column for a Galactic dust-to-gas ratio
of $\nh/A_V=1.8\times10^{21}$~cm$^{-2}$ \citep{predehl95a}. The
one-component fits, on the other hand, for the most part fall below
the column density expected from the observed color by factors of
$\ga$10. These fits are primarily in the low-SNR outskirts of the
absorbing region (Figure \ref{fig:ncomp}), and favor a
$\tau_{5890}\sim0.1$, $C_f\sim1.0$ fit to \nad\ vs. the
$\tau_{5890}\sim2$, $C_f\sim0.2$ fit that emerges with better SNR and
two components. The observed emission line optical depths in these
same regions (\S\,\ref{sec:obs}) are also consistent with the
optically-thick solution. The logical inference is that the
one-component fits underestimate the true column density in the
absorbing gas, and that higher SNR at these locations would yield the
correct optical depth and covering factor ($\tau_{5890}\sim2$ and
$C_f\sim0.2$). However, emission line filling could also affect the
fits in these spaxels (\S\,\ref{sec:disc_structure}).

\begin{figure*}[t]
  \centering \includegraphics[width=7in]{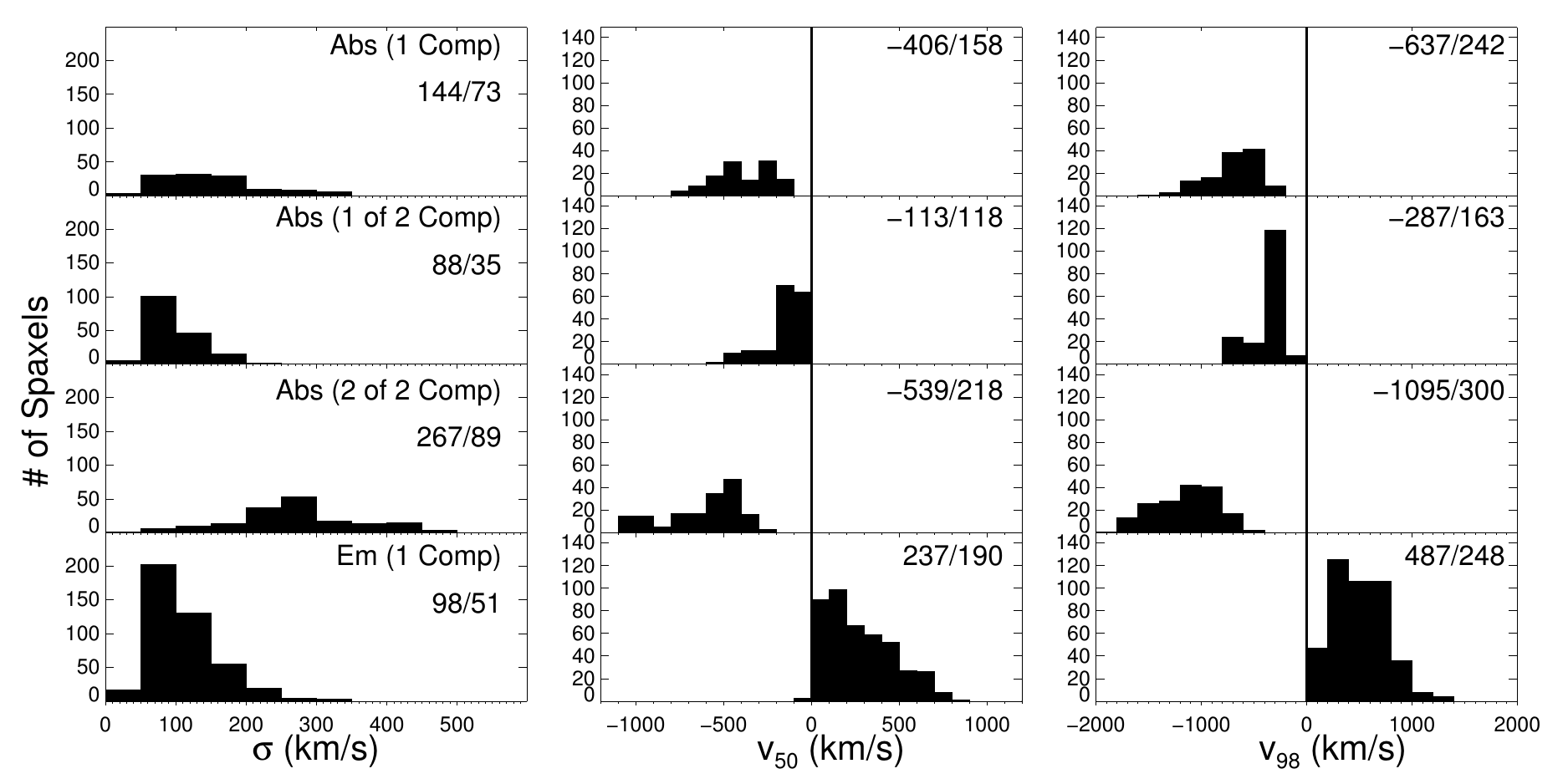}
  \caption{Histograms of velocity dispersion (left), $\vfifty$
    (middle), and $\vtsig \equiv \vfifty \pm 2\sigma$ (right) in fits
    to \nad. Absorption lines fits with one (top) or two (middle)
    components are shown, as well as emission line fits (bottom). The
    median and standard deviation of each distribution are listed in
    the upper right corner of each panel. The absolute values of the
    peak emission line velocities are smaller than those of the
    absorption lines.}
  \label{fig:velhist}
\end{figure*}

\begin{figure}[b]
  \centering \includegraphics[width=3in]{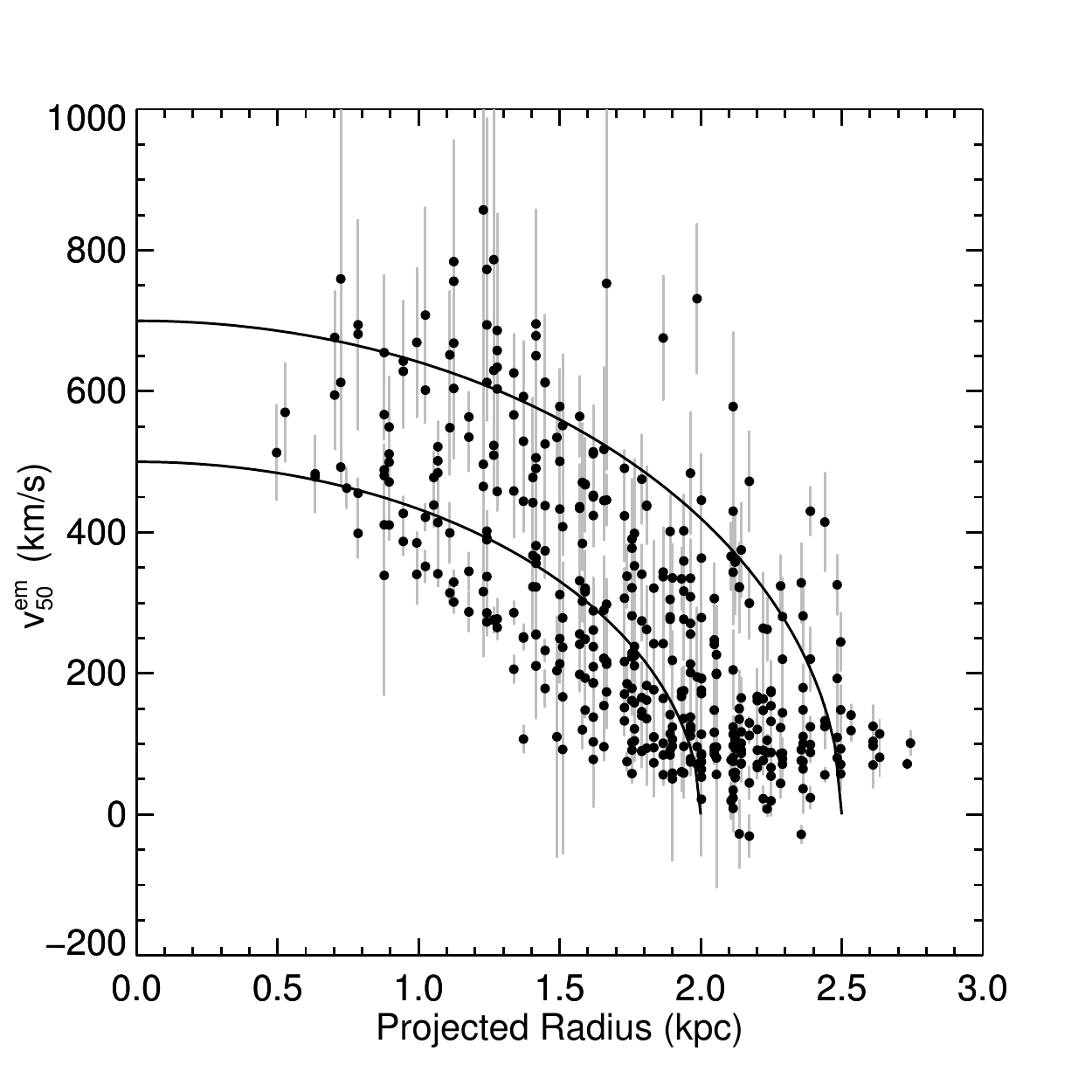}
  \caption{\nad\ emission line velocity (\vfifty) vs. projected
    galactocentric radius. The lines show projected velocities of thin
    shells with $v = 500$ \kms, $R = 2.0$ kpc (bottom line) and
    $v = 700$ \kms, $R = 2.5$ kpc (top line). Though thin shell models
    are consistent with the data, the dependence of velocity on radius
    may result in part from emission line filling of absorption lines
    (\S\,\ref{sec:disc_velocity}).}
  \label{fig:vel_v_r}
\end{figure}

This physical correspondence between the column densities of neutral,
outflowing gas and optical continuum attenuation has several possible
consequences, all of which can be inferred from the current data but
which need further confirmation with other data to be more generally
applicable. (1) The apparent \weqa-color correlation in the absorbing
gas is deceptive (Figure \ref{fig:weq_v_col}). The physical
correlation observed in the two-component fits in Figure
\ref{fig:nh_v_col} is more fundamental and less obvious. (2)
Correlations between \weqa\ and galaxy color have been previously used
to argue that the neutral, outflowing gas probed by \nad\ in dusty
active galaxies is also dusty
\citep{veilleux95a,heckman00a,shih10a,rupke13a}. However, these
conclusions need to be revisited and better quantified using high SNR
measurements of \nad. (3) Accurate measurements of \nh\ in these
galaxies requires $\text{SNR}\ga15$ in the equivalent width
(Figure~\ref{fig:ncomp}). (4) The \nad\ column densities reported from
fits to high SNR nuclear spectra are reasonably accurate
\citep[e.g.,][]{heckman00a,rupke05b,rupke05c,rupke11a,rupke13a}. This
implies furthermore that the Galactic conditions used to calculate
\nh\ \citep{rupke05a} are close to the conditions in the cool,
foreground gas of dusty outflows in active galaxies. (5) The
filamentary continuum structures observed in optical \hst\ images of
ULIRGs are intimately connected with the outflows in these
systems. This idea is explored further in
\S\,\ref{sec:disc_structure}.

If the dusty outflow attenuates the stellar continuum, it should also
attenuate the \nad\ emission line flux, as suggested by the strong
correlation between the two (Figure~\ref{fig:weq_v_col}). It will not,
however, affect the emission line equivalent width if it is a screen
in front of both the line emission and stars, since it will attenuate
both equally. This is consistent with the lack of correlation between
$F435W-F814W$ and $W_{eq}^{em}$. In \S\,\ref{sec:disc_structure}, a
possible model for the \nad\ line emission from the wind is presented.

\subsection{Outflow Velocities} \label{sec:disc_velocity}

The absorption-line velocities in nuclear spectra of ULIRGs show high
blueshifts with a detection rate of 80\%, which has been interpreted
as evidence for ubiquitous outflowing gas
\citep{heckman00a,rupke02a,rupke05b,rupke05c,martin05a}. These neutral
outflows have higher velocities in galaxies hosting strong AGN
\citep{rupke13a}, an effect that is also seen in the molecular
\citep{veilleux13a,cicone14a} and ionized
\citep{westmoquette12a,rupke13a,arribas14a} gas phases. Since
F05189$-$2524 is a ULIRG powered by a powerful quasar, high-velocity
outflowing gas in the neutral gas phase should be observed. However,
the highest projected outflow velocities are typically observed
outside of the nuclear line of sight \citep{rupke13a}.

The velocities of resonant line emission due to outflows in ULIRGs are
unknown. Observations of two high-redshift galaxies with strong,
extended \ion{Mg}{2} 2796, 2800 \AA\ emission both reveal emission
lines slightly less offset from systemic than the corresponding
absorption lines. \citet{rubin11a} outline a case with mean absorption
velocities of $-200$ to $-300$ \kms\ and flux-weighted emission
velocities of 30$-$70 \kms, while \citet{martin13a} present a galaxy
with a velocity of $-230$ \kms\ in absorption and $110$~\kms\ in
emission (though in this study the absorption line velocity was
derived from rest-frame UV \ion{Fe}{2} lines). Models predict smaller
absolute velocity shifts of emission lines than absorption lines in a
given resonant transition \citep{prochaska11a}. For
spherically-symmetric outflow models, the emission line profile is
created by multiple absorptions and emissions throughout the nebula,
and the integrated emission more closely reflects the average dynamics
of the entire outflow. However, it is still redshifted because bluer
photons are absorbed by the outflowing regions on the near side of the
galaxy \citep{prochaska11a}.

To parameterize the neutral gas velocities using the fits, each
component was considered separately and \vfifty\ (the center of the
analytic profile) and $\vtsig = \vfifty\pm2\sigma$ ($+$ for emission,
$-$ for absorption) were calculated. The empirical method for
constraining \nad\ also yields velocity information, and for the
absorbing gas this broadly matches the fit results. However, the
empirical emission line velocities are noisy due to the low SNR. The
fits can also decompose the profile properly into the two doublet
components. The empirical measurements do not distinguish between the
doublet components except on the profile edges, and thus introduce a
velocity uncertainty.

Figures~\ref{fig:velmaps} and \ref{fig:velhist} display maps and
histograms of the spatially-resolved velocity distributions in
F05189$-$2524.

The absorption line velocities are everywhere blueshifted, and in both
components they peak in a ridge running 1 kpc E of the nucleus to 1
kpc S of the nucleus, coincident with the peak in the total equivalent
width. The linewidth in the broad component reaches
$\sigma = 500$~\kms, leading to peak velocities of almost
$-$2000~\kms. These velocities are higher than any measured to date in
the neutral phase of an extended outflow in nearby ULIRGs, including
Mrk~231 \citep{rupke05b,rupke05c,rupke11a,rupke13a}, though ionized
gas velocities can exceed this. For spaxels with two fitted
components, median values of \vfifty\ (\vtsig) are $-$110 and $-$540
\kms\ ($-$300 and $-$1000~\kms). These values are slightly larger than
earlier single-aperture measurements due to the latter being
luminosity-weighted. \citet{rupke05c} find two absorption line
components in the nuclear spectrum of F05189$-$2524 (through a
1\farcs0 slit), with \vfifty\ (\vtsig) values of $-$90 and
$-$400~\kms\ ($-$220 and $-$900~\kms). \citet{teng13a} find the same
velocity components in deep \ion{H}{1} data (\vfifty $=$ $-$140 and
$-$440~\kms, with their data adjusted to correspond to the $z_{sys}$
used in this paper). The high neutral gas velocities are consistent
with those measured in other major mergers containing a quasar
\citep{rupke13a}.

The emission line velocities have a strong spatial dependence. The
line widths do not vary strongly across the FOV, but the values of
\vfifty\ and \vtsig\ fall with increasing radius (Figure
\ref{fig:vel_v_r}). The highest velocities (exceeding 1000~\kms) are
not detected at the edges of the FOV, but only within $1-2$~kpc of the
nucleus. Peak redshifted velocities in the emission lines are smaller
than peak blueshifted velocities in the absorption lines, with \vtsig\
reaching only 1300~\kms\ in emission (Figure~\ref{fig:velhist}).

\begin{figure}[t]
  \centering \includegraphics[width=3.5in]{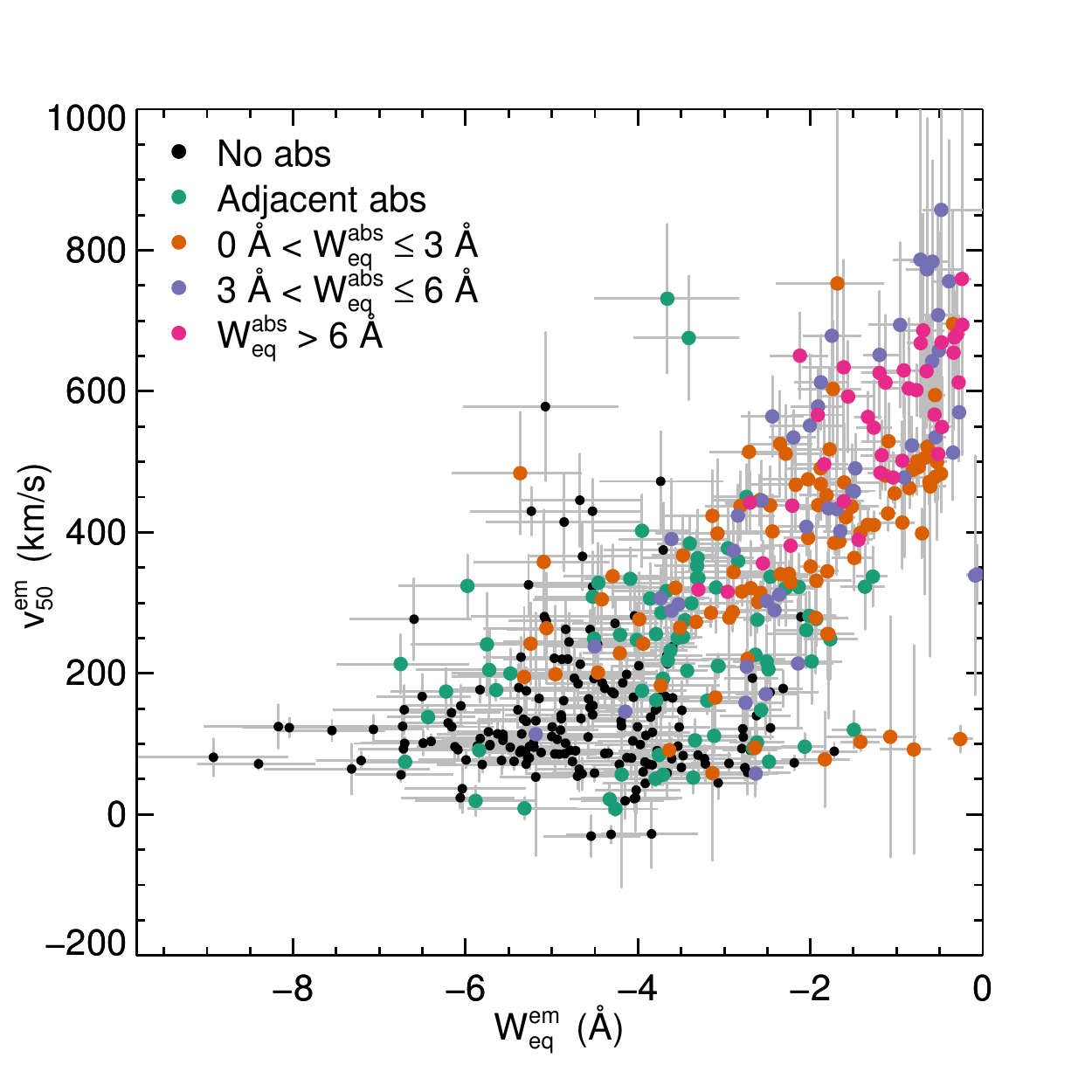}
  \caption{\nad\ emission line velocity (\vfifty) vs. \nad\ emission
    line equivalent width. The two are strongly correlated, and are
    also correlated with the strength of the \nad\ absorption in the
    same spaxel (as indicated by the colors). Two models that could
    explain this are (1) a spherically symmetric shell whose projected
    velocity is higher at smaller radii (Figure \ref{fig:vel_v_r}) or
    (2) 3D emission-line filling of absorption lines at low equivalent
    widths, which preferentially fills in velocities near systemic
    (\S\,\ref{sec:disc_velocity} and \,\ref{sec:disc_structure}).}
  \label{fig:emvel_v_emweq}
\end{figure}

The spatial distribution of emission line \vfifty\ and \weqe\ are very
similar. In fact, the two are strongly correlated
(Figure~\ref{fig:emvel_v_emweq}). In Figure~\ref{fig:emvel_v_emweq},
the spaxels are colored according to whether or not they are adjacent
to absorption lines (if \nad\ absorption is not present in a spaxel)
and according to the absorption line equivalent width (if \nad\
absorption is present in a spaxel). It is apparent that on average,
higher \weqa\ corresponds to lower $|\weqe|$ and higher emission line
velocity. Furthermore, even when absorption is not formally fitted,
there is a trend for spaxels with neighboring absorption line fits to
have smaller $|\weqe|$.

There are two interpretations of these correlations, both of which
relate to wind structure (which is discussed further in
\S\,\ref{sec:disc_structure}).

First, Figures~\ref{fig:vel_v_r} and \ref{fig:emvel_v_emweq} could
reflect the intrinsic spatial variation of the projected emission line
velocities. If the outflow is an expanding thin shell of gas, then the
resonant emission from the far side of the outflow will have a
projected velocity that decreases with increasing projected
radius. The correlation of velocity with equivalent width in this
scenario then implies that $|\weqe|$ decreases with increasing
projected radius. In Figure~\ref{fig:vel_v_r} two lines show the
result of geometric projection of two thin shells, one with $v=500$
\kms\ and $R=2.0$~kpc and the other with $v=700$ \kms\ and
$R=2.5$~kpc. This range of radii and velocities is consistent with the
bulk of absorbing gas on the near side.

A second interpretation is that the increase of redshift with
decreasing equivalent width is due to emission line filling of nearby
(in wavelength space) absorption, as predicted by
\citet{prochaska11a}. In this model, the neutral gas density can vary
smoothly with (actual) radius, rather than being concentrated in a
thin shell.  Filling of the line absorption by emission scattered into
the line of sight in a given spaxel simultaneously decreases \weqa\
and $|\weqe|$. It should also reduce \sbe\ and increase \vfifty, since
the line filling of a spherical wind (at least in integrated spectra)
occurs preferentially near the systemic velocity
\citep{prochaska11a}. The fact that velocity also correlates with
\weqa\ can be explained in this model if the emission line equivalent
width is most affected when the absorption is strongest (i.e., there
is more absorption to fill in).

\section{DISCUSSION} \label{sec:discussion}

The present data are only the second spatially-resolved detection of
resonant line emission from cool gas in a galactic wind at low
redshift (and the first 3D observations at any redshift). These data
thus uniquely constrain current models of cool gas in galactic winds
\citep{prochaska11a}, as well as serving as a benchmark for future
observations and models.

\subsection{Clues to the Wind Structure from Resonant Emission and
  Absorption} \label{sec:disc_structure}

\citet{prochaska11a} presented the first radiative transfer models of
cool gas emission and absorption in simple, spherically-symmetric
galactic winds that extend from the nucleus out to large radii. They
persuasively argued that both absorption and emission are strong in
dust-free spherical winds, that the absorption lines are blueshifted
and the emission lines near systemic or redshifted, and that the
emission extends over the entire wind area with a relatively shallow
but declining radial surface brightness profile.

This is broadly consistent with our data. In particular, we observe
strong, blueshifted absorption and redshifted emission, as well as
emission that extends across the FOV. However, the emission is
strongly suppressed in the nuclear regions.

\begin{figure}[t]
  \centering \includegraphics[width=3.5in]{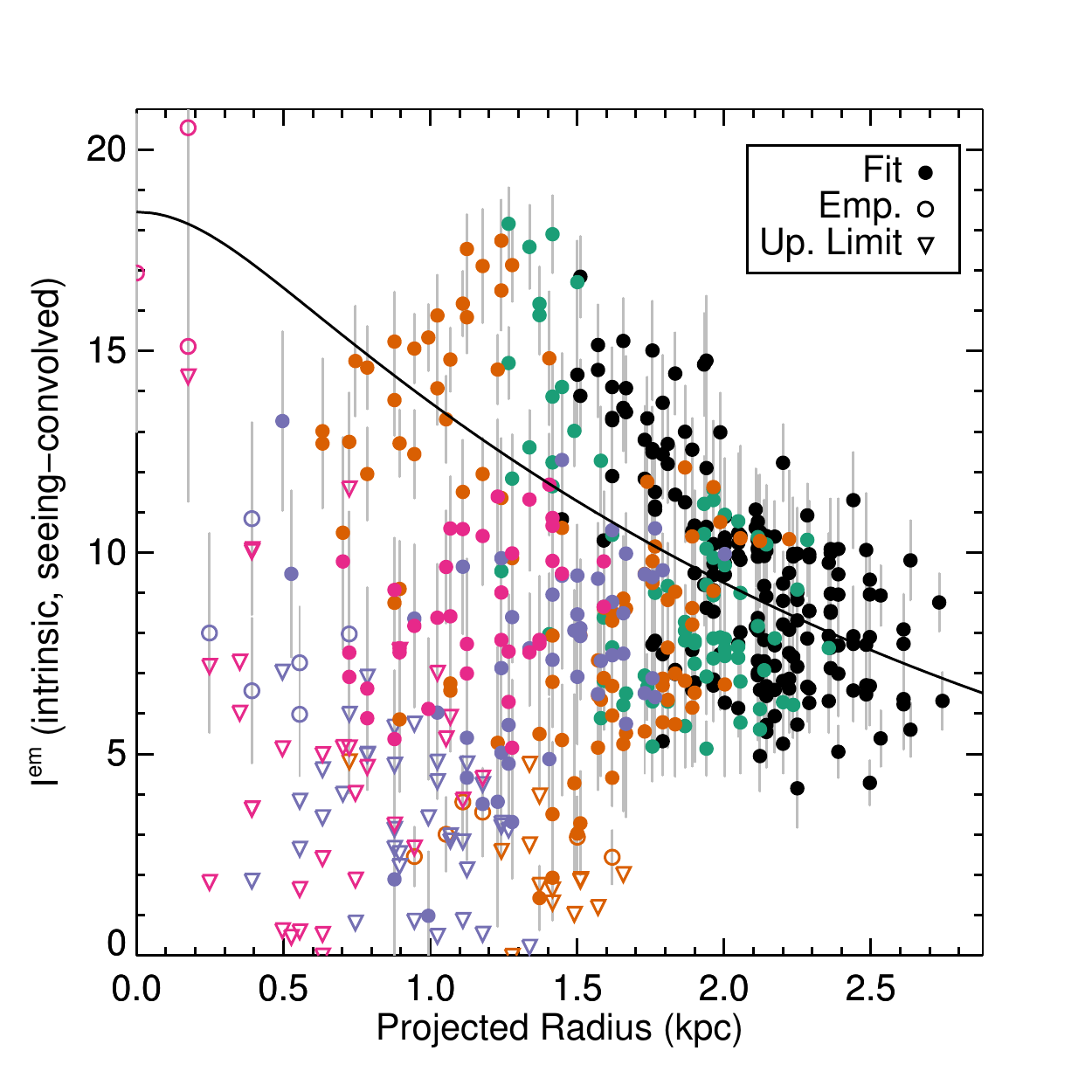}
  \caption{Intrinsic \nad\ emission line surface brightness
    vs. projected radius. Measured values of \sbe\ have been corrected
    for intrinsic obscuration using continuum colors
    (\S\,\ref{sec:weq_flux}). The solid points show data from fits to
    the emission lines, and the open points and triangles show
    empirical measurements and upper limits in the inner regions where
    fits are not available. The points are colored according to the
    contamination from \nad\ absorption (see
    Figure~\ref{fig:emvel_v_emweq} for colors). The line is the best
    fit model to the data, which is a \sersic\ profile with
    $R_e = 2.5$~kpc and $n = 1.0$. We fit only points with emission
    line fits and $\weqa < 3$~\AA\ to minimize contamination from
    emission line filling (\S\,\ref{sec:disc_structure} and Figure
    \ref{fig:model}).}
  \label{fig:sb_v_rad}
\end{figure}

\begin{figure*}
    \centering \includegraphics[width=7in]{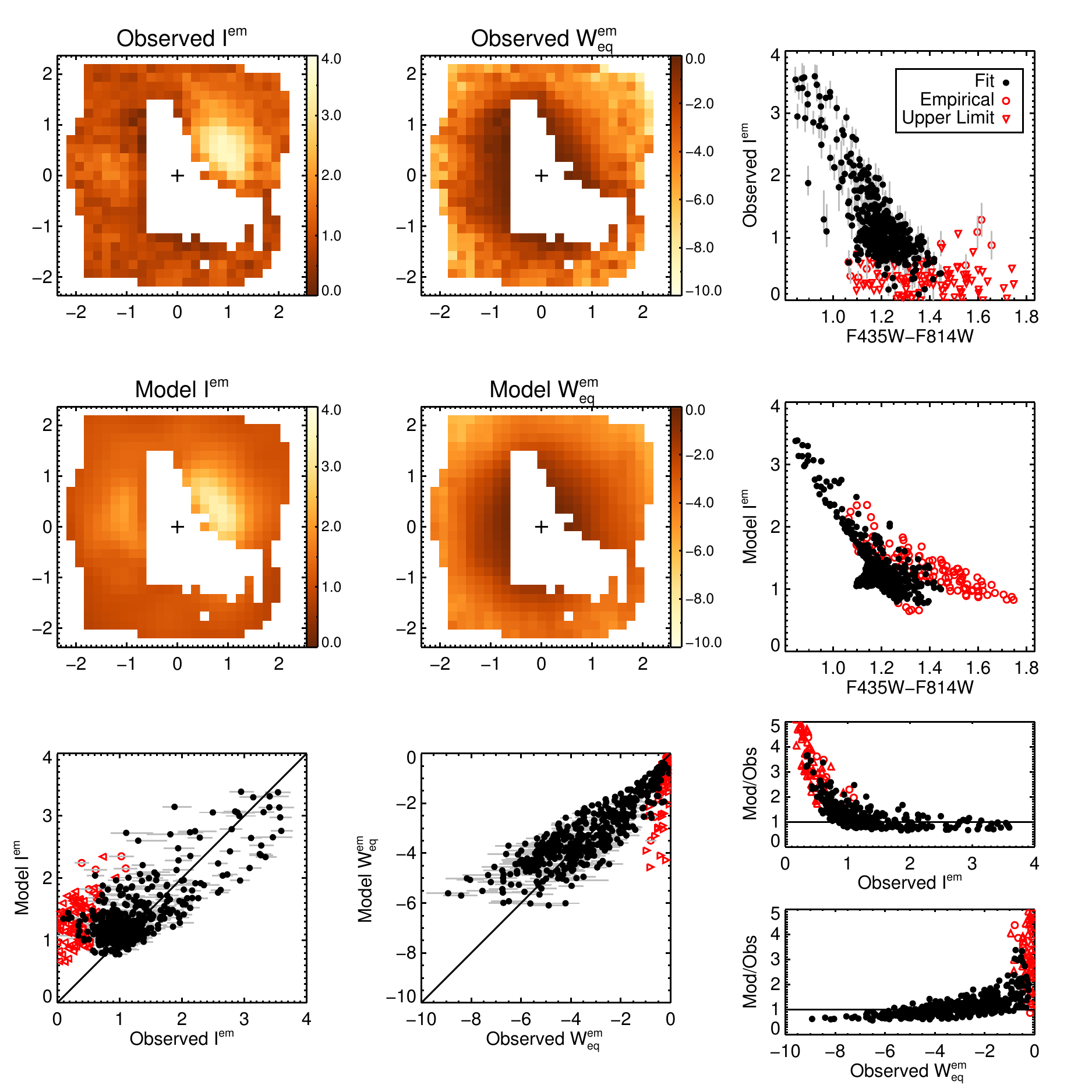}
    \caption{\scriptsize Comparison between observed emission line
      surface brightnesses, equivalent widths, and continuum colors
      (top row) and those predicted by the dust-extincted model
      (middle row). The maps in the middle row show model spaxels for
      which there is a fitted emission line. The far right middle
      panel shows these same model points as black circles, while
      model points which correspond to spaxels in which only empirical
      estimates or upper limits exist are shown as red open circles.
      The bottom row quantitatively compares the observed and model
      values for \sbe\ and \weqe. The best fit model is a \sersic\
      profile with $n=1.0$, $I_e=2.1\times10^{-15}$ erg s$^{-1}$
      cm$^{-2}$ arcsec$^{-2}$, and a scale length $R_e=2.5$ kpc. This
      profile is extincted by a foreground screen whose magnitude is
      inferred from the stellar continuum colors and convolved with
      the seeing. There is excellent overall correspondence between
      the two (bottom row) except at the lowest values of \sbe\ and
      \weqe\ (bottom right). At these levels, the spaxels are
      primarily in the inner kpc in regions of strong absorption and
      no emission line fit; these spaxels are represented in the above
      plots by empirical estimates and upper limits. Some emission
      line filling of the absorption lines, which reduces both \sbe\
      and \weqe, may explain the discrepancy
      (\S\,\ref{sec:disc_structure}).}
  \label{fig:model}
\end{figure*}

To further explore an emission line model that fills the FOV,
Figures~\ref{fig:sb_v_rad} and \ref{fig:model} compare the emission
line data to a model assuming azimuthal symmetry in the plane of the
sky. The model uses a \sersic\ surface brightness profile for the
integrated \nad\ emission, $I(R) = I_e e^{-(R/R_e)^{(1/n)}}$. This
profile was extincted by a foreground screen whose magnitude is
inferred from the stellar continuum colors and convolved with the
seeing. Only points with low absorption-line contamination, to
minimize emission line filling, were fit. The attenuation curve
assumed was \citet{calzetti00a} with $R_V = 4.85$ (the curve closest
to the data in Figure \ref{fig:nh_v_col}), though the result is not at
all sensitive to $R_V$. The background continuum for computing \weq\
was the average observed optical continuum in the 50~\AA\ above and
below \nad. A best-fit model with $I_e = 2.1\times10^{-15}$ erg
s$^{-1}$ cm$^{-2}$ arcsec$^{-2}$, $R_e = 2.5$~kpc, and $n=1.0$
reproduces the ridge of enhanced emission W of the nucleus, the
general features of the equivalent width map, and the correlation of
continuum colors with \sbe. This is similar to the $R$ and $H$-band
stellar surface brightness profiles, which have $n = 0.8-1$ and
$R_e = 3.9-4.3$ \citep{veilleux02a,veilleux06a}.

At the lowest values of \sbe\ and \weqe, including the central region
where only empirical upper (lower) limits exist for \sbe\ (\weqe) and
\nad\ absorption is the strongest, the model begins to deviate from
the data (bottom panels of Figure \ref{fig:model}). This range of
\weqe\ also corresponds to the largest deviations of \vfifty\ from
systemic (Figure~\ref{fig:emvel_v_emweq}). It is likely that in these
regions the emission and absorption lines are being affected by
emission-line filling of the absorption lines
\citep{prochaska11a}. Besides reducing \weqe\ and increasing \vfifty\
for the emission lines, this effect will also reduce \weqa\ in these
regions by up to $1-2$ \AA, and may push the absorption line solution
towards lower optical depths since the filling will preferentially
affect the lower optical depth \nad\ line, 5896~\AA. Figure
\ref{fig:nh_v_col} suggests that the latter effect is small for most
two component absorption line fits, but could impact the low SNR, one
component fits.

However, emission line filling is not the only factor responsible for
the projected radial velocity profile of the emission line gas
(Figure~\ref{fig:vel_v_r}). It is apparent that the maximum redshifts
(\vtsig) in the emission lines decrease away from the nucleus
(Figure~\ref{fig:velmaps}), implying that projection effects are also
reducing the measured velocities.

This 2D view of the wind does not directly constrain whether the wind
is configured as a thin shell or rather as a radially-continuous,
volume-filling sphere. At face value, a radially thin shell in
emission would be limb-brightened, contrary to the data, while a
radially-filling wind would have the qualitative surface brightness
behavior that is observed \citep{prochaska11a}. However, radiative
transfer effects can impact the naive expectation, and intermediate
cases such as a thicker shell are also plausible. Further detailed
modeling is required before quantitative surface brightness profiles
like that discussed here can be used to constrain the density profile
of the wind.

It is evident from this analysis that resonant line emission and
absorption are complementary probes of galactic outflows. The \nad\
absorption in this system traces the dusty outflowing gas
($A_V = 1-4$, $\nh\sim10^{21-22}$) that happens to lie along the line
of sight to the galaxy nucleus and its stellar disk. The \nad\
emission, conversely, is almost invisible along these same lines of
sight because of dust obscuration. Along unobscured lines of sight,
emission provides a view of the rest of the wind. This latter view is
complicated by radiative transfer effects which lower the measured
velocity if the wind is volume-filling rather than configured as a
shell \citep{prochaska11a}.

The presence of \nad\ emission across the FOV suggests a wind that is
more extended than one would infer from \nad\ absorption alone (and
vice versa, since the two peak in strength in very different
places). Absorption line probes show that the neutral phase of the
outflows can extend to 10~kpc or farther in ULIRGs
\citep{martin06a}. The wind in this system extends to at least the
edge of the FOV (3~kpc), and is apparently optically thick to these
radii in integrated line emission (\S\,\ref{sec:obs}). Deep long-slit
observations may constrain the point at which the wind becomes
optically thin, and thus the true size of the wind. This probe will be
more sensitive to the true extent than absorption lines, since no
background continuum is required and the equivalent width is
increasing with increasing radius (Figure~\ref{fig:map_fitweq}). The
two high-$z$ examples also show a more extended outflow in emission
than absorption \citep{rubin11a,martin13a}.

The collimation of the F05189$-$2524 outflow is consistent with that
of other ULIRG winds, though the current data do not constrain the
inner collimation. Spatially-resolved studies of neutral, ionized, and
molecular gas outflows in ULIRGs reveal that the gas preferentially
emerges along the minor axis of the nuclear disk at radii of
$1-2$~kpc, but becomes less collimated at larger radii
\citep{rupke05b,rupke13a,rupke13b}. If an observer could choose a
random nuclear line of sight through a given wind, they would detect
an outflow in absorption $\sim$80\%\ of the time \citep{rupke05b}. The
absorption line map of this galaxy is consistent with this
result. Although we detect nuclear absorption, our observations trace
only one nuclear sightline through the wind. A clumpy and/or
collimated wind could be less detectable through other nuclear
sightlines. The data are consistent with a collimated inner wind if
the nuclear disk is near face-on or an uncollimated wind if the
nuclear disk is closer to edge-on (there is no published data on the
nuclear disk in this galaxy). Our model for the observed \nad\
emission is symmetric in the plane of the sky, but since it is
detected only at scales $\ga$0.5~kpc this is consistent with either an
uncollimated wind or the decreasing collimation with increasing radius
in other ULIRGs outflows \citep{rupke13a}.

These results also show that foreground dust is one of the primary
drivers of the observed morphology of \nad\ emission and absorption in
this galaxy. Na atoms ionize easily (the first ionization potential is
5.1~eV), so the spatial correspondence between absorbing Na atoms and
dust is natural, since dust can shield the atoms from ionizing
radiation. The observed peak in \nad\ flux
(Figure~\ref{fig:map_fitweq}) is deceptive, in that it more closely
reflects the foreground screen than the actual distribution of
emission line gas (which in the model is azimuthally symmetric).


\subsection{Detection of Extended \ion{Na}{1}~D
  Emission} \label{sec:det}

As is evident from many long-slit studies \citep{heckman00a, rupke02a,
  rupke05a, rupke05b, martin05a}, high surface brightness resonant
line emission is not a common feature of galactic winds in nearby
galaxies. \citet{prochaska11a} list dust, wind asymmetry, and
emission-line filling of absorption lines as possible physical causes
of these non-detections.

\nad\ line emission is in fact present at low levels in stacked
spectra of face-on galaxies with low dust attenuation \citep{chen10a},
suggesting that both dust and asymmetry could indeed play a role in
affecting detectability of \nad\ emission. At high $z$,
\citet{kornei13a} found that $A_{UV}$ is at least partly responsible
for the varying strength of \ion{Mg}{2} absorption and \ion{Fe}{2}$^*$
emission in galactic winds. The equivalent width and/or detection rate
of \ion{Mg}{2} emission in high $z$ star forming galaxies is also
higher in galaxies with bluer UV colors
\citep{weiner09a,erb12a,kornei13a}.

The current data show that dust is pivotal in shaping the morphologies
of both \nad\ absorption and emission. However, it does not obviously
affect the detectability of \nad\ emission. The $A_V$ values in
F05189$-$2524 are typical for a ULIRG
\citep{surace98a,veilleux99a}. The highest emission line flux is
indeed in the region of lowest extinction and absent in the regions of
highest extinction, but is detectable throughout the FOV, up to 3 kpc
from the galaxy nucleus. As for asymmetry, the 3D structure of the
outflow in F05189$-$2524 is not constrained beyond the surface
brightness emission model, but the data do not point to any unusual
asymmetry that would affect the resonant lines differently than in
other systems.

The two high $z$ galaxies with extended \ion{Mg}{2} emission have not
only blue UV colors, but also high rest-frame blue luminosity compared
to similar galaxies \citep{rubin11a,martin13a}. A necessary physical
condition for strong resonant lines is enough continuum emission at
the same wavelength as the transition to excite the atoms out of the
ground state. Higher $R$-band luminosity could thus in principle
strengthen \nad\ emission.

In fact, within a nuclear aperture of 4 kpc, F05189$-$2524 has
$M_{R,4\text{kpc}} = -20.44$, which is brighter than 90\%\ of ULIRGs
in the 1~Jy sample \citep{kim02a} and is negligibly affected by the
nuclear point source \citep{kim13a}. Furthermore, most of the galaxies
that are brighter than F05189$-$2524 are Seyfert 1s, in which the
light is concentrated in an unresolved point source and possibly
beamed preferentially along the line of sight to Earth. Excluding the
Seyfert 1s in the parent sample of ULIRGs, F05189$-$2524 is brighter
than 95\%\ of the parent sample at $R$, within a 4~kpc nuclear
aperture. Of the five non-Seyfert-1 ULIRGs that have lower
$M_{R,4\text{kpc}}$ than F05189$-$2524, four have been previously
observed with long-slit spectra \citep{rupke05a,rupke05b,rupke05c} and
one (Mrk~273) also with IFS \citep{rupke13a}. None show obvious \nad\
emission.

This suggests that the nuclear $R$-band luminosity of F05189$-$2524 is
responsible for the detection of extended \nad\ emission, though it
may not in fact be a sufficient condition for its detection. However,
deeper IFS observations of other dusty, $R$-luminous systems will be
necessary to confirm this. In long-slit observations, slit loss could
be also a factor if the line emission is spatially extended at a low
surface brightness \citep{prochaska11a}, though that is clearly not a
factor in the current data or in other IFS studies of \nad\
\citep{rupke13a}. Nonetheless, it could be responsible for the lack of
observed \nad\ emission in long slit data
\citep{rupke05a,rupke05b,rupke05c,martin05a}. Furthermore, purely
nuclear spectra in these dusty systems are less likely to show \nad\
emission, since the nucleus is typically the most heavily obscured
part of the galaxy and will be dominated by absorption rather than
emission.

Several other nearby galaxies show resonant emission from metal lines,
but there is presently nothing to distinguish their luminosities. NGC
1808 is fainter at $R$ (in terms of integrated brightness) than other
nearby AGN \citep{koss11a}, with $M_R\sim-21$ (NED). In a near-UV
study of five $z\sim0.25$ ULIRGs, \citet{martin09a} find \ion{Mg}{2}
emission in two Seyfert 2s, but in neither case is this emission
extended. The relevant continuum luminosity for this transition at
these redshifts is observed-frame $U$-band. No $U$ measurements exist
for this sample, though integrated $U$-band luminosities in nearby
ULIRGs do not reveal significant differences between Seyfert 2s and
non-Seyfert spectral types \citep{surace00a}.

In other respects, however, the \nad\ absorption and emission in NGC
1808 are similar to that described here. NGC 1808 has a nuclear
obcuration similar to F05189$-$2524 ($A_V = 3-4$;
\citealt{kewley01b}). Its outflowing dust streamers, which are
prominent in optical color maps ($B-R$), also show \nad\
absorption. However, these streamers are less obscured than much of
the \nad\ absorbing region of F05189$-$2524 ($A_V < 1$;
\citealt{phillips93a}). Finally, the emission on the far side of NGC
1808 emerges in a relatively unobscured region; they appear just
outside dust lanes in the galaxy disk.

Should \nad\ emission be detectable outside the GMOS FOV? The emission
line model for F05189$-$2524 has an exponential scale length of
2.5~kpc, and spaxels near the field edge (at $1-2$ scale lengths)
still have $\sbe \sim 10^{-16}$ erg s$^{-1}$ cm$^{-2}$
arcsec$^{-2}$. \citet{prochaska11a} predict that line emission extends
to a radius where the optical depth still exceeds a few tenths, while
the \nad\ line emission in the GMOS FOV is optically
thick. Furthermore, if the wind is isotropic, the total absorption
line and emission line equivalent widths of an integrated spectrum
should be equal \citep{prochaska11a}. However, summing the GMOS
spectra across the FOV yields $\weqa = 3.1$~\AA\ and
$\weqe = -0.7$~\AA. If the wind in F05189$-$2524 is fairly isotropic,
a significant amount of emission (in terms of equivalent width) lies
outside the GMOS FOV. A deep long-slit spectrum or narrow-band image
should be able to constrain the extent of the \nad\ emission in this
system.

How would F05189$-$2524 look at a greater distance? Almost all
observations of resonant line emission have been made at $z \ga 0.5$
with rest-frame UV lines. At these redshifts, \nad\ is in the
observed-frame near-infrared, a wavelength range in which sensitive
absorption line measurements of individual objects are challenging due
to sky emission and absorption lines and lower CCD
sensitivity. Furthermore, in this system the primary source of Poisson
noise is not from the emission lines but rather the continuum.

Despite these difficulties, detection of \nad\ emisison in an object
similar to F05189$-$2524 is reasonable at $z\la0.5$ if strong sky
lines are avoided. For instance, at $z=0.3$ a 1\farcs0 slit covers the
GMOS FOV. Summing the GMOS spectra across the FOV in the present data
shows that emission is detected with $\weqe = -0.7$~\AA\ and a flux of
$1.5\times10^{-15}$~erg s$^{-1}$ cm$^{-2}$. Cosmological surface
brightness dimming at $z\sim0.3$ amounts to a factor of three. The
Gemini/GMOS Integration Time Calculator indicates that less than an
hour of spectroscopic integration in a 1\arcsec$\times$1\arcsec\
aperture yields a signal-to-noise ratio in which the emission can be
easily distinguished from the continuum (the emission line flux is
$\sim$10$\times$ the noise from the continuum). Even at $z=0.5$, the
continuum signal-to-noise ratio in one hour is high enough that the
emission line flux is several times larger than the noise in the
continuum. At higher redshifts, the detectability drops sharply due to
sky lines and CCD sensitivity.

\section{SUMMARY} \label{sec:summary}

We present the second example of resolved \nad\ emission from a
galactic wind in the nearby universe. The GMOS IFU follows \nad\
absorption and emission across a $4\times4$~kpc aperture in a nearby
ULIRG hosting a quasar. These resolved observations provide unique
constraints on models and future observations of resonant line
emission in galactic winds.

The \nad\ absorption lines trace the dusty near side of the
wind. F05189$-$2524 shows the highest velocity kpc-scale neutral gas
outflow to be directly resolved in any nearby galaxy, with blueshifted
velocities up to almost $-$2000~\kms. The present IFS data directly
connect the absorption in the outflow with the foreground dust column
obscuring the stellar continuum. Fits to the absorption lines with low
covering factors yield optically thick absorption with
$\nh\sim10^{21-22}$~cm$^{-2}$. The column density is correlated with
the dust column in a way that matches the column density expected if
the wind has a Galactic dust-to-gas ratio.

The optically thick \nad\ emission lines extend to the edges of the
FOV, but are extincted in the inner regions of the galaxy. An
azimuthally symmetric \sersic\ model, with the emission extincted by
the same foreground screen as the stellar continuum, reproduces the
observed emission line surface brightness and equivalent width maps at
all but the lowest values of \sbe\ and $|\weqe|$. At these levels,
which correspond to the inner regions of the wind nearest the regions
of \nad\ absorption, emission line filling of the \nad\ absorption
lines near the systemic velocity is a likely culprit, as suggested by
correlations of \weqe\ with emission line velocity and \weqa.

The morphology of the resonant line absorption and emission in this
galaxy are clearly regulated by the presence of dust. Furthermore, the
detection of extended \nad\ emission in this system in particular may
be due to its unusually high continuum surface brightness at the rest
wavelength of the resonant transition, similar to galaxies with
extended \ion{Mg}{2} emission observed in the rest-frame UV at high
$z$ \citep{rubin11a,martin13a}. Future observations of \nad\ in other
nearby or low redshift systems ($z \sim 0-0.5$) with high rest-frame
$R$-band surface brightness may be able to detect extended \nad\
emission due to galactic winds.

This unique dataset shows that \nad\ absorption and emission are very
complementary probes of wind structure. \weqa\ and \weqe\ are
anti-correlated on a spaxel-by-spaxel basis, and the \sbe\ and \weqa\
peaks are on opposite sides of the nucleus. The absorption lines trace
the near side of the wind along the line of sight to Earth, while the
emission lines trace the bulk of the interior and far side of the wind
through radiative transfer \citep{prochaska11a}. Together, they form a
picture of a dusty, wide angle wind that extends to at least several
kpc from the nucleus in most directions. This picture is consistent
with previous work on galactic winds in major mergers
\citep{rupke05b,martin06a,rupke13a}.

\acknowledgments We thank the referee for their thorough and helpful
comments. This work was based on observations obtained at the Gemini
Observatory (program ID GS-2011B-Q-64), which is operated by the
Association of Universities for Research in Astronomy, Inc., under a
cooperative agreement with the NSF on behalf of the Gemini
partnership: the National Science Foundation (United States), the
Science and Technology Facilities Council (United Kingdom), the
National Research Council (Canada), CONICYT (Chile), the Australian
Research Council (Australia), Minist\'{e}rio da Ci\^{e}ncia,
Tecnologia e Inova\c{c}\~{a}o (Brazil) and Ministerio de Ciencia,
Tecnolog\'{i}a e Innovaci\'{o}n Productiva (Argentina). D.S.N.R. was
supported by a Cottrell College Science Award from the Research
Corporation for Science Advancement and by NASA grant Keck/JPL RSA
1461849. S.V. was supported in part by NSF grant AST1009583 and NASA
grants NHSC/JPL RSA 1427277 and 1454738.

The \hst\ observations described here were obtained from the Hubble
Legacy Archive, which is a collaboration between the Space Telescope
Science Institute (STScI/NASA), the Space Telescope European
Coordinating Facility (ST-ECF/ESA) and the Canadian Astronomy Data
Centre (CADC/NRC/CSA).


\end{document}